\documentclass[reprint,superscriptaddress, nofootinbib,longbibliography]{revtex4-2}
\pdfoutput=1
\usepackage{graphicx}
\usepackage{amssymb}
\usepackage[linktoc=none]{hyperref}
\usepackage[dvipsnames]{xcolor}
\usepackage{dcolumn}
\usepackage[english]{babel}
\usepackage[utf8x]{inputenc}
\usepackage[T1]{fontenc}
\usepackage{eqnarray,amsmath,physics}
\usepackage{appendix}
\usepackage{xcolor}

\usepackage{hyperref}
    \hypersetup{
    colorlinks   = true,    
    urlcolor     = blue,    
    linkcolor    = blue,    
    citecolor    = red      
    }

\DeclareMathAlphabet{\mathpzc}{OT1}{pzc}{m}{it}

\newcommand{\vek}{\Vec{k}}

\newcommand{\INFN}{INFN - Sezione di Napoli, Complesso Univ. Monte S. Angelo, I-80126 Napoli, Italy}
\newcommand{\UNINA}{Physics Department "Ettore Pancini", Universit\'a degli studi di Napoli "Federico II", Complesso Univ. Monte S. Angelo, Via Cintia, I-80126 Napoli, Italy}
\newcommand{\UNISA}{Physics Department "E.R. Caianiello", Universit\'a degli studi di Salerno, Via Giovanni Paolo II, 132, I-84084 Fisciano (Sa), Italy}
\newcommand{\CNR}{CNR-SPIN Napoli Unit, Complesso Univ. Monte S. Angelo, Via Cintia, I-80126 Napoli, Italy}

\begin{document}
\title{Strain-induced topological phase transition at (111) SrTiO$_3$-based heterostructures}

\author{M. Trama}
\email{mtrama@unisa.it}
\affiliation{\UNISA}
\affiliation{\INFN}

\author{V. Cataudella}
\email{cataudella@na.infn.it}
\affiliation{\UNINA}
\affiliation{\CNR}

\author{C. A. Perroni}
\email{carmine.perroni@unina.it}
\affiliation{\UNINA}
\affiliation{\CNR}
\begin{abstract}
    The quasi-two-dimensional electronic gas at the (111) SrTiO$_3$-based heterostructure interfaces is described by a multi-band tight-binding model providing electronic bands in agreement at low energies with photoemission experiments. We analyze both the roles of the spin-orbit coupling and of the trigonal crystal field effects. We point out the presence of a regime with sizable strain where the band structure exhibits a Dirac cone whose features are consistent with \textit{ab-initio} approaches. The combined effect of spin-orbit coupling and trigonal strain gives rise to non-trivial spin and orbital angular momenta patterns in the Brillouin zone and to quantum spin Hall effect by opening a gap at the Dirac cone. The system can switch from a conducting to a topological insulating state \textit{via} modification of trigonal strain within a parameter range which is estimated to be experimentally achievable.  
\end{abstract}
\maketitle
\section{Introduction}
The discovery \cite{ohtomo2004high} of a quasi-two-dimensional electronic gas (q2DEG) formed at the (001) interface between the SrTiO$_3$ (STO) and  LaAlO$_3$ (LAO) grown on it gave rise to a very rich research field. 
For example, the q2DEGs formed in these oxide heterostructures are characterized by the simultaneous presence of a strong spin-orbit coupling (SOC) \cite{caviglia2010tunable} and superconductivity \cite{reyren2007superconducting}. 
Both these phenomena are tunable by electric field effect \cite{caviglia2010tunable,reyren2007superconducting,caviglia2008electric}. Superconductivity can coexist with a 2D-magnetism \cite{hwang2012emergent,pai2018physics} in q2DEGs with suitable atomic engineering of LAO/STO interfaces \cite{stornaiuolo2016tunable}.
Furthermore, q2DEGs in (001) LAO/STO interfaces have been theoretically proposed to show topological properties in normal \cite{vivek_normal} and superconducting phases of 2D \cite{scheurer2015topological,mohanta2014topological,loder2015route,fukaya2018interorbital} and quasi-1D models \cite{fidkowski2011majorana,fidkowski2013magnetic,mazziotti2018majorana,perroni2019evolution,perroni_ultimo}.
It is therefore not surprising that oxides heterostructures, in particular LAO/STO interfaces, have been proposed for the realization of low dimensional devices in the field of spintronics and quantum information \cite{bibes2007oxide,bibes2011ultrathin,dey2019oxides,chakraborty2020perovskite,massarotti2020high}.

A general property of heterostructures, among which LAO/STO, is the possibility of controlling strain at the interfaces by realizing superlattices or using micromembranes \cite{wong2009superlattice, xie2018coherent, sambri2020self}. The presence of strain allows the manifestation of phenomena absent in the bulk structure \cite{perroni2003modeling, iorio2011electron, biswas2017strain, deng2019strain}. Moreover, these phenomena can be tuned by manipulating the strain itself. In particular,  the possibility of strain-tuned topological phase transition has been theoretically predicted \cite{ochi_strain}  and experimentally realized \cite{mutch_strain}. 


After the discovery of the presence of the q2DEGs on interfaces with different crystallographic cut directions \cite{herranz2012high}, the attention was extended to the (111) LAO/STO interface, in which even more exotic phenomena have been predicted \cite{boudjada2017magnetic}. Recently anisotropic magnetoresistance both in normal \cite{rout2017six} and superconducting state \cite{ monteiro2017two, davis2017magnetoresistance} has been observed, suggesting the unconventional nature of superconductivity host by (111) LAO/STO interface.
Further, recent studies \cite{doennig2013massive} have discovered a striking similarity between this interface and graphene due to the presence of a Dirac point in the band structure. This similarity arises by virtue of the similar honeycomb lattice structure formed at the interface when STO is Ti-terminated.
\\The analogy between graphene and the q2DEG in the (111) interface suggests the possibility of non-trivial topological effects \cite{chakhalian2020strongly}. 
In fact, since the Haldane seminal discovery \cite{haldane1988model}, it has been known that q2DEG on honeycomb lattices in an insulating state can manifest Quantum Hall Effect (QHE) in the absence of magnetic field. The subsequent discovery of the Quantum Spin Hall Effect (QSHE) by Kane and Mele \cite{kane2005z,kane2005quantum} showed that topologically protected edge states can appear in graphene even in presence of time reversal invariance, when SOC is included. 
This suggested that topological insulators (TIs) can be achieved in materials with strong SOC \cite{bernevig2006quantum,moore2007topological,fu2007topological}: this has been later experimentally confirmed \cite{konig2014quantum,hsieh2008topological,xia2009observation}. Further studies have highlighted the presence of TI in 3D bulk \cite{Topo_3D} and their nanostructures \cite{Topo_perroni1,Topo_perroni2}.
Finally, Xiao \textit{et al.} \cite{xiao2011interface} proposed the use of Transition Metal Oxides (TMOs) based heterostructures, in particular perovskites, in the (111) direction for realizing TIs. This possibility is offered by the presence of strong atomic SOC typical of the TMOs and the peculiar geometry of (111) perovskites. The authors used Tight Binding (TB) methodologies to show the presence of a QSHE in different materials. This methodology has been widely used to compute theoretically the electronic band structure for LAO/STO (111) \cite{monteiro2019band,khanna2019symmetry,boudjada2017magnetic}.
\\In this work we analyze specifically the properties of the $(111)$ STO interface electronic structure using an effective TB model, focusing on the LAO/STO one. Compared to the previous works which adopted TB methodologies we emphasize the role of the strain at interface and the atomic SOC in the band structure properties. 
Such an approach is useful to have a physical understanding of the mechanisms giving rise to the peculiar features of the system.
We perform a comparison with the experimental data available on the (111) interface. In particular we fix the hopping parameters from the Angle-Resolved Photoemission Spectroscopy (ARPES) experiments on the bare (111) STO surface.
We identify the region of the parameter space in which the Dirac point emerges within the electronic band structure; we find that this happens when a relevant term in the TB treatment is the strain. As already mentioned, the presence of a Dirac Point in the electronic band structure of the (111) LAO/STO interface was already predicted in \cite{doennig2013massive}, where Density Functional Theory (DFT) was adopted. The use of TB methodology allows an effective modeling of the band structure near the Dirac point including also the effects of the SOC with a small computational effort and, then, open up the way to study the transport properties and the topological invariants of this interface that can be very demanding for "first principle" approaches.

We verify the concrete possibility of the regime with sizable strain studying the relation between the energy parameter in the TB model and the geometry of the physical structure. Furthermore we include an analysis of the SOC role in the properties of the energy spectra. We also investigate the presence of spin and orbital angular momentum patterns in the Brillouin Zone (BZ).
In the last part of the work we perform a systematic scan of the parameter space to identify the region in which the system exhibits a QSHE. We verify the bulk-boundary correspondence by explicitly showing the presence of edge states on finite-thick zigzag ribbons. Our results suggest the possibility of a strain-induced topological transition for the system.

This paper is organized as follows: in Section \ref{model} the model Hamiltonian for the STO 2D bilayer is introduced, discussing every contribution we take into account. The methods used to estimate the hopping parameters is also presented. In Section \ref{TrigonalPart} the properties of the energy spectra are explored, emphasizing the role of the trigonal strain; we also discuss the physical range in which the strain is expected to vary. Furthermore the presence of spin and angular momentum patterns in the BZ is shown. In Section \ref{topological} the topological properties of the system are discussed together with the methods used to determine the topological invariant $Z_2$. In Section \ref{conclusioni} we report our conclusions.

\section{Model and Methodology}\label{model}
\begin{figure}[t!]
    \includegraphics[width=0.45\textwidth]{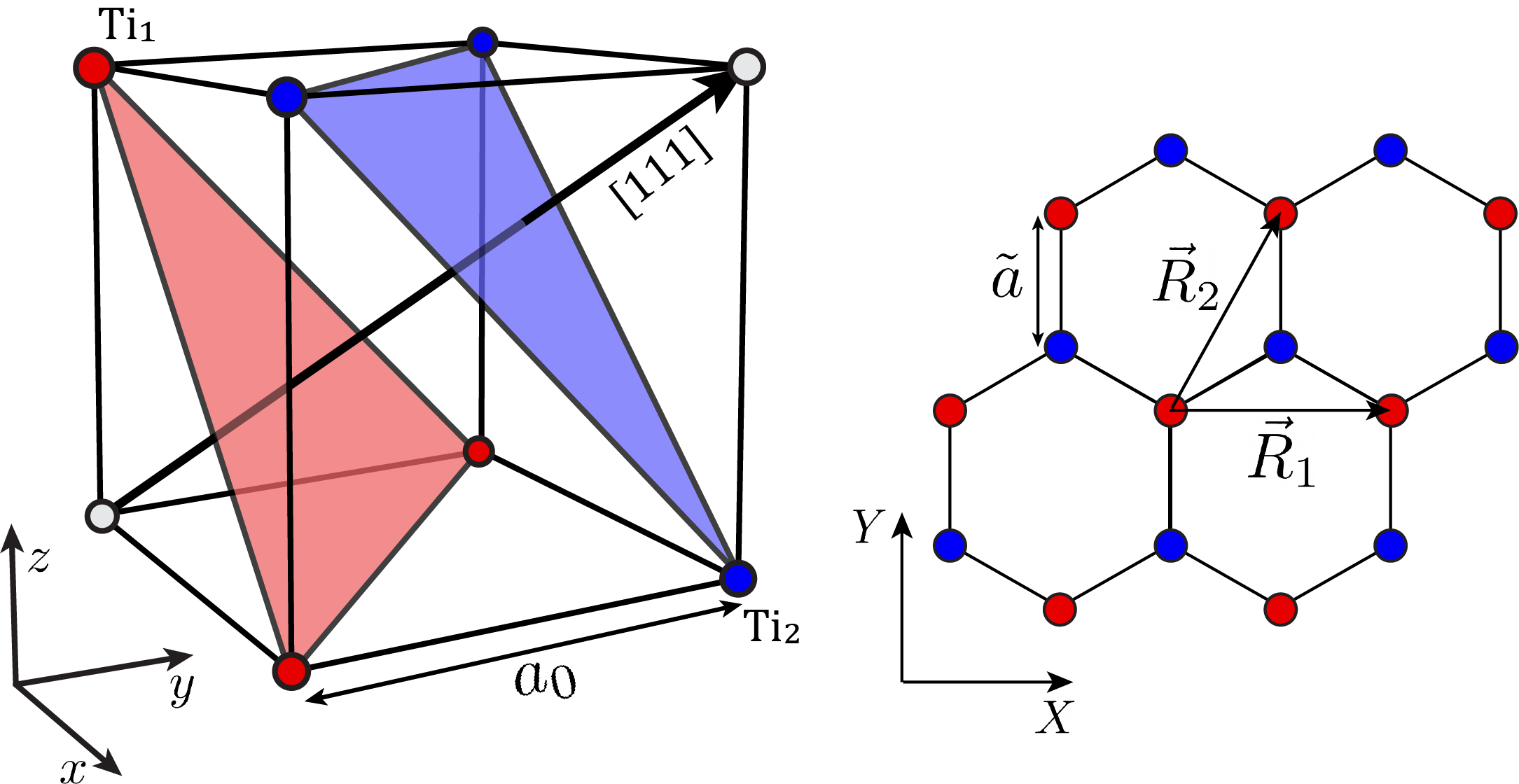}
    \caption{(Left) Schematic representation of the Ti atoms in STO lattice. The red and blue dots represent two different non-equivalent atoms belonging to the two different planes depicted in the figure. (Right) Projection on the (111) plane of the Ti atoms with $\tilde{a}=a_0\sqrt{2/3}$; $\Vec{R}_1$ and $\Vec{R}_2$ identify our choice of the primitive vectors of the lattice.}
    \label{fig:Honey}
\end{figure}
In this section we discuss the electronic band structure of the q2DEG using a minimal TB approach. This methodology allows us to encode the physical properties of the crystal in a small number of energy parameters in order to understand their role in the band structure formation.
\\The projection of two different Ti atom layers along the (111) direction of the STO results in a honeycomb structure, similar to the graphene one, with a lattice constant of $\tilde a=0.3905\sqrt{2/3}$ nm as shown in Fig.~(\ref{fig:Honey}).
The conduction band is obtained by the superposition of the 3d orbitals of Ti atoms of the STO. Due to the approximate cubic geometry these orbitals split up into a triplet $t_{2g}=\{d_{yx},d_{zx},d_{xy}\}$ and a doublet $e_g=\{d_{z^2},d_{x^2-y^2}\}$ of orbitals, as described in Ref. \cite{Keppler1998}. The $t_{2g}$ orbitals are responsible for the formation of the lower conduction band, so that in the following we will only consider them in the description of the energy structure.
We consider the two dimensional structure originating from the projection on a single plane of only two layers of Ti, whose atoms are labeled using the symbols Ti$_1$ and Ti$_2$. We denote by  $d_{i\alpha\sigma,\vek}$ the annihilation operator of the electron of 2D dimensionless quasi-momenta $\Vec{k}=\tilde{a}\Vec{K}$, where $\Vec{K}$ is the quasi-momentum, occupying the orbital $i=\{xy,yz,zx\}$ belonging to the layer $\alpha=\{\text{Ti}_1,\text{Ti}_2\}$ and of spin $\sigma=\pm 1/2$. With this notation we studied the following Hamiltonian
\begin{equation}
    H=H_{TB}+H_{TRI}+H_{SOC}+H_{v}.
\end{equation}
We will discuss the different contributions separately. We start from the TB Hamiltonian which in the $k$-space takes the form
\begin{equation}
    H_{TB}=\sum_{\vek}\sum_{i,\alpha\beta,\sigma}t_i^{\alpha\beta}(\Vec{k}) d_{i\alpha\sigma,\vek}^{\dagger} d_{i\beta\sigma,\vek}
    \label{TBk},
\end{equation}
where $t_i^{\alpha\beta}(\Vec{k})$, considering also the next-to-nearest neighbour (NNN) hopping, contain three contributions:
\begin{itemize}
    \item the first and most important contribution will be the head-on overlap of two lobes of the same kind of orbitals for nearest neighbours. This happens between a Ti$_1$ and a Ti$_2$, so that this term will describe an interlayer hopping. The amplitude for this hopping will be denoted by $t_3$;
    \item the second contribution will come from the lateral overlap of two lobes of two orbitals of the same type between Ti$_1$ and Ti$_2$. The amplitude of this interlayer term is denoted by $t_2$;
    \item the third and smallest term will be a NNN intralayer hopping caused by the head-on overlap between two orbitals of two Ti on the same layer. The amplitude of this term will be denoted as $t_1$. From best fit calculations this term will be neglected, as we can show in the following.
\end{itemize}
The matrix representation of the Hamiltonian~(\ref{TBk}) is given in Appendix \ref{appendiceLikelihood} (Eq.~(\ref{TBmatrix})).
The values of the hopping parameters have been fixed by comparison with ARPES experiments \cite{walker2014control} to the values
$
        t_3=0.5\hspace{0.2cm}\rm{eV};
$
$        t_2=0.04\hspace{0.2cm} \rm{eV};
$
$
        t_1=0\hspace{0.2cm} \rm{eV},
$ obtained as best fit parameters of a maximum likelihood analysis which is detailed in the Appendix \ref{appendiceLikelihood}.
\\$H_{TRI}$ takes into account the possibility of having strain at the interface along the (111) direction. The physical origin of this strain is the possible contraction or dilatation of the crystalline planes along the (111) direction. The strain at the interface can be manipulating \textit{via} deposition of other crystal, such as LAO. This coupling is taken into account by the following contribution, whose form is fixed by symmetry considerations \cite{khomskii2014transition}
\begin{equation}
    H_{TRI}=\frac{\Delta}{2}\sum_{\vek}\sum_{i\neq j,\alpha,\sigma} d_{i\alpha\sigma,\vek}^{\dagger} d_{j\alpha\sigma,\vek}.
    \label{eq:trigonal}
\end{equation}
In the following we will refer to this term as the trigonal crystal field. More details about this term will be given in the next section. We initially point out that the sign of $\Delta$ determines the kind of the applied strain along the (111) direction: if $\Delta<0$ a contraction of the Ti atom planes occurs, while for $\Delta>0$ a dilatation occurs.
\\Another contribution taken into account is the atomic SOC, which is represented in the $k$-space by
\begin{equation}
    H_{SO}=\frac{\lambda}{2}\sum_{\vek}\sum_{ijk,\alpha,\sigma\sigma'}i\varepsilon_{ijk}
    d_{i\alpha\sigma,\vek}^{\dagger} \sigma^{k}_{\sigma\sigma'}d_{j\alpha\sigma',\vek}
    \label{eq:spinorbit}
\end{equation}
where $\varepsilon$ is the Levi-Civita tensor, and $\sigma^k$ are the Pauli matrices. 
Finally we include a possible interaction with an external electric field orthogonal to the planes of the form
\begin{equation}
    H_v=\frac{v}{2}\sum_{\vek}\sum_{i,\alpha,\sigma}\xi_{\alpha} d_{i\alpha\sigma,\vek}^{\dagger} d_{i\alpha\sigma,\vek},
    \label{electric}
\end{equation}
where $\xi_{Ti_1}=+1$ and $\xi_{Ti_2}=-1$. 
This contribution will take into account the possibility of a small perturbation at the interface breaking the inversion symmetry between the two layers.

\section{Trigonal Crystal Field effects}\label{TrigonalPart}
\begin{figure}[h!]
   \includegraphics[width=0.45\textwidth]{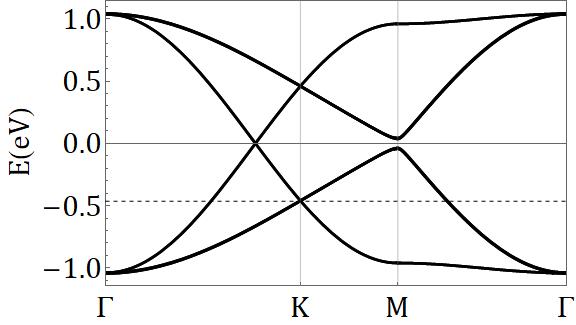}\\
   \hspace{0.2cm}
   \includegraphics[width=0.425\textwidth]{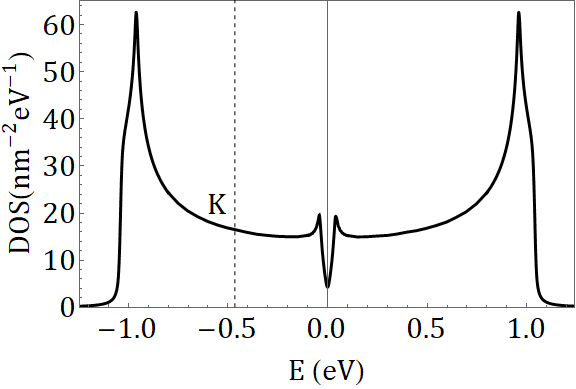}
    \caption{(Upper panel) Energy band structure and (Lower panel) Density of states for $\Delta=-0.005$ eV, $\lambda=0$ eV and $v=0$ eV. The energy $E=-0.46$ eV corresponding to the states at the $K$ point is highlighted by a dashed line.}
    \label{fig:BandDSmallNoLNov}
\end{figure}
\begin{figure*}
    \centering
    \includegraphics[width=0.43\textwidth]{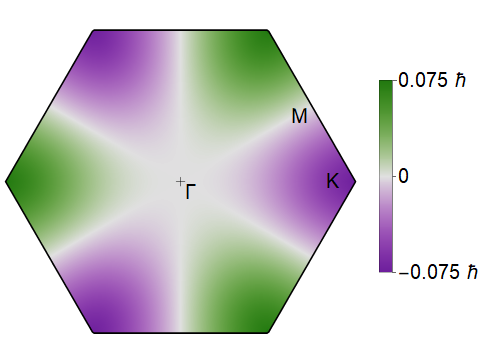}
    \hfill
    \includegraphics[width=0.40\textwidth]{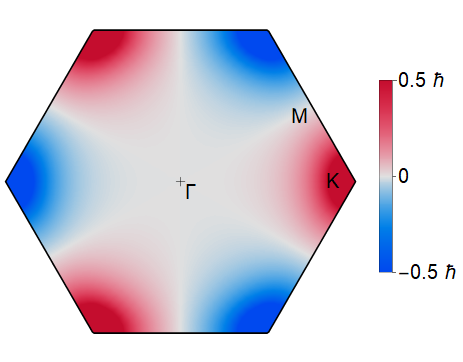}\\
    \caption{Mean values of the out-of-plane components of (left panel) orbital and (right panel) spin angular momentum, localized on one of the two Ti layers, for $\Delta=-1$ eV, $\lambda=0.1$ eV and $v=0$ eV and with $\mu=-1.14$ eV (within the gap at the Dirac point).}
    \label{L111S111}
\end{figure*}

The purpose of this section is to illustrate the role of the trigonal crystal field in determining the properties of system. We will show two extreme situations: a first one in which $\Delta$ is a small perturbation compared to the TB and a second one in which $\Delta$ is the dominant parameter.
\begin{figure}[h!]
    \centering
    \includegraphics[width=0.45\textwidth]{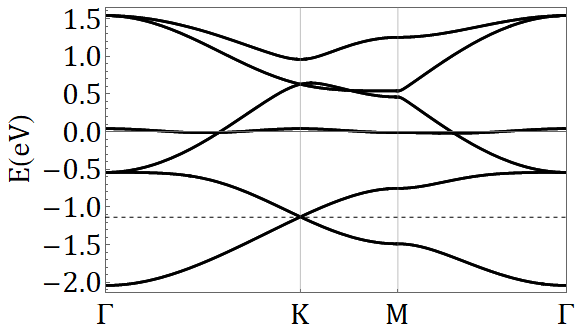}\\
    \includegraphics[width=0.44\textwidth]{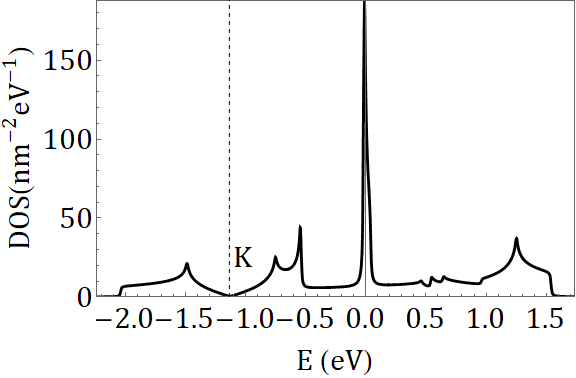}
    \caption{(Upper panel) Energy band structure and (Lower panel) Density of states for $\Delta=-1$ eV, $\lambda=0$ eV and $v=0$ eV. The energy $E=-1.14$ eV corresponding to the Dirac point is highlighted by a dashed line.}
    \label{fig:d-1}
\end{figure}
In order to exemplify the behaviour of the system for small values of $\Delta$, we choose a benchmark $\Delta=-0.005$ eV, equal to the trigonal coupling for the unstrained interface \cite{de2018symmetry}. For this choice the energy band structure and the density of states are shown in Fig.~(\ref{fig:BandDSmallNoLNov}). At the $K$ point of the Brillouin zone one can notice an energy dispersion which is linear in lower bands: this could erroneously suggest that a Dirac cone is present within the energy bands. In Appendix \ref{unstrained section} we discuss a more complete characterization of the energy spectra near the $K$ point, finding that the bands do not have the typical conical structure of the Dirac point in this region. 
\\\\The band structure drastically changes when the parameter $\Delta$ is assumed to be larger, exhibiting a Dirac cone in the low energy region as shown in Fig.~(\ref{fig:d-1}). 
In fact when $|\Delta/t_3|\gtrsim1$ there is a splitting of the energy levels (neglecting the spin degeneracy) in a pair of bands of lower energy and four bands of higher energy. In the limit of $|\Delta/t_3|\gg1$ these bands asymptotically become the eigenstates of the trigonal Hamiltonian called respectively $a_g$ and $e_g^\pi$ \cite{khomskii2014transition}, belonging to the corresponding symmetry group. The hopping contribution causes only a weak mixing between the two groups of states and gives rise to a Dirac cone at the $K$ point in the Brillouin zone between the $a_g$ states. Our results are in agreement with the results obtained using DFT methodologies in Ref.~\cite{doennig2013massive} We can locally adopt a perturbative approach near the K point to obtain an effective Dirac Hamiltonian
\begin{equation}
    H_{{\rm{eff}}}=\hbar v_F \hspace{0.1cm} \Vec{\delta k}\cdot \Vec{\sigma} \otimes \tau_z +\Delta_{SO}\sigma_Z\otimes\tau_z+\rm{const},
    \label{diracHam}
\end{equation}
from which we can extract the Fermi velocity of the electron at the Dirac point, which is $v_F=(\frac{1}{2}(t_2+2t_3))/\hbar\approx2.5\times10^{5}$ m s$^{-1}$, and $\Delta_{SO}$ is the effective coupling induced by the SOC. It depends on the parameters as $\Delta_{SO}\sim(\lambda t_3^2)/\Delta^2$, since it originates from the perturbation to the third order in the TB and SOC contributions. 
A more detailed discussion is provided in Appendix \ref{appendice LARGE DELTA}. The Dirac point is sensitive to the SOC, which opens a gap between the bands; this is described by the $\Delta_{SO}$ term in the Hamiltonian~(\ref{diracHam}).
\\The two dimensional carrier density $n_{2D}$ needed to fill the bands up to the Dirac point is of the order of $10^{14}$ cm$^{-2}$ (for $\Delta=-1$ eV, $n_{2D}=7.6\times10^{14}$ cm$^{-2}$, corresponding to 0.34 electrons for each Ti ion). This is comparable with the typical electron densities injected at the LAO/STO interface, making it possible to perform an experimental investigation via field effect technique.
\\\\The presence of the trigonal distortion induces some peculiar behaviour of the angular and spin polarization between the two layers of Ti atoms. In order to show this we evaluated for a fixed filling $\mu$ the following quantity:
\begin{equation}
    \langle\mathcal{O}_{\alpha}\rangle(\Vec{k})=\sum_{E<\mu}\bra{E,\vek}P_{\alpha} \mathcal{O}\ket{E,\vek}
    \label{operator}
\end{equation}
where $\mathcal{O}_\alpha$ is the generic component of the angular momentum or spin momentum projected on the $\alpha$-th layer, $\ket{E,k}$ is a generic eigenstate of energy $E$ and quasi-momentum $\Vec{k}$ and $P_\alpha$ is a projection operator on the states localized on the $\alpha$-th layer
\begin{equation}
    P_\alpha=\sum_{i,\sigma,\vek} \ket{\psi_{i\alpha\sigma,\vek}}\bra{\psi_{i\alpha\sigma,\vek}},
\end{equation}
where we defined $\{|\psi_{i\alpha\sigma,\Vec{k}}\rangle\}$ as the vector of the original basis of the Hamiltonian.
These quantities evaluated throughout the Brillouin zone are depicted in Fig.~(\ref{L111S111}) for a benchmark choice of the parameters.
This Figure shows that a magnetic moment localized in $\vek$ is formed between the two layers, in principle observable via spin-polarized ARPES experiments. A peculiar feature appears in Fig.~(\ref{L111S111}): when the chemical potential is within the gap formed at the Dirac point, the out-of plane spin component is totally localized near the $K$ point in the Brillouin zone, while the orbital angular momentum is almost zero. The latter can be understood by observing that in the limit $-\Delta/t_3\gg1$ the lower orbitals tend to the $a_g$ states which are characterized by $\langle \Vec{L}\rangle=0$. The localization of the spin, instead, can be understood by observing that far from the Dirac point the splitting among the first and second doublet is induced by TB alone, and therefore the mean value of the spin vanishes. On the other hand nearby the Dirac point Eq.~(\ref{diracHam}) shows that the splitting induced by TB vanishes. The gap at the Dirac point is induced by the SOC, represented by the $\Delta_{SO}$ term in Eq.~(\ref{diracHam}), leading to a non vanishing mean value of the spin on each of the two layers oriented along the (111) direction. Of course, in the absence of inversion symmetry, e.g. $v=0$ in our case, the sum over the two layers of the projected mean values vanishes. This is reasonable and in agreement with literature \cite{bhowal2021orbital,xiao2010berry}. Furthermore, since the system is time reversal invariant in the absence of a magnetic field, it does not show magnetic behaviour and a many body interaction is needed in order to recover ferromagnetic effects: for this reason, the total spin integrated over the whole Brillouin zone is always vanishing.
\subsection{Connection between trigonal coupling and physical strain}
\begin{figure}
    \centering
    \includegraphics[width=0.35\textwidth]{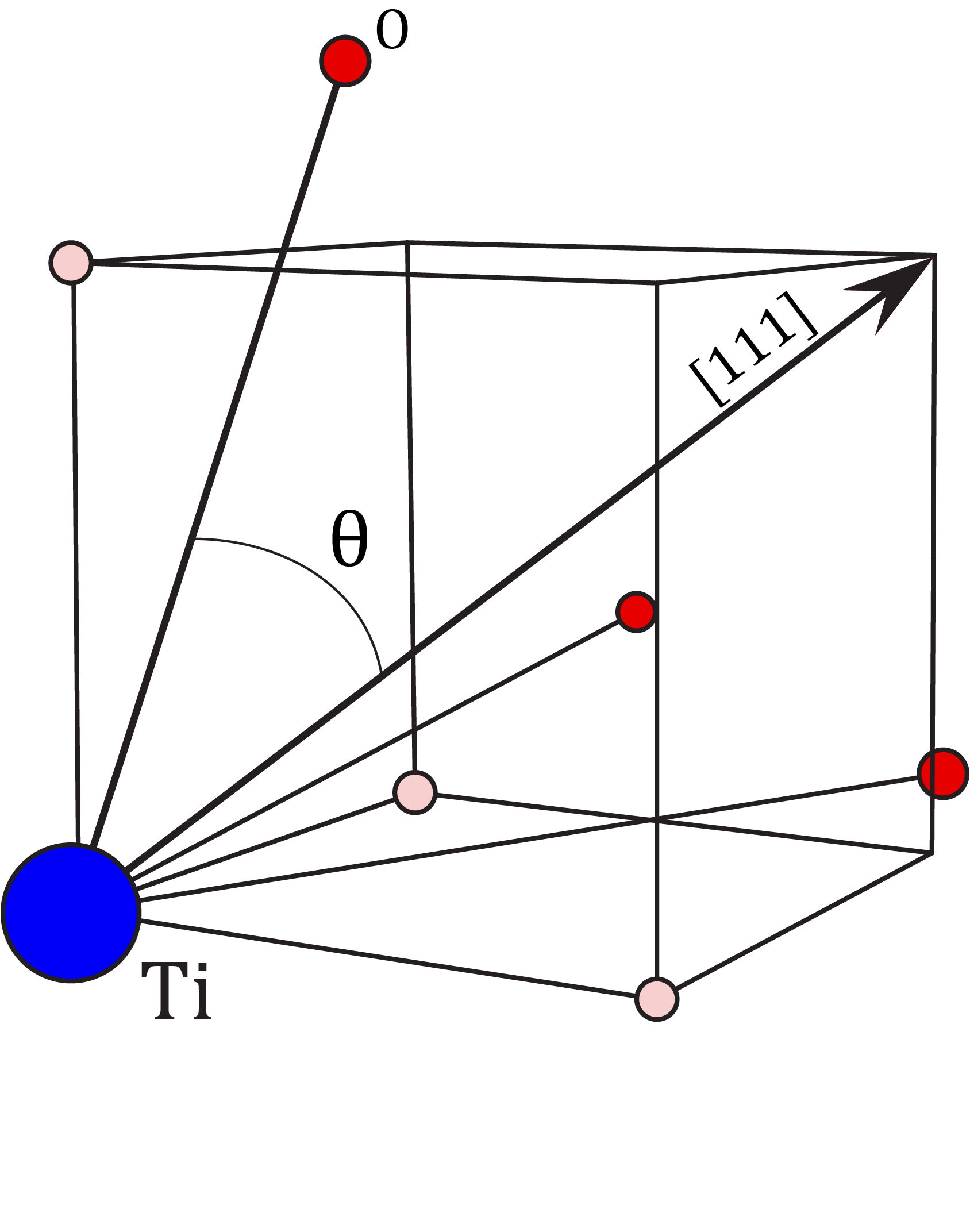}
    \caption{Cubic structure under a trigonal distortion. The blue sphere is the Ti atom, while the red ones are the O atoms. The pink spheres are the positions of the O atoms in the undistorted structure. The black arrow is the (111) direction. $\theta$ represents the distortion angle used for the parametrization of the strain: when $\theta=\arccos{(1/\sqrt{3})}$ the structure is unstrained; smaller (larger) values of $\theta$ lead to dilatation (contraction) along (111) direction.}
    \label{fig:cage}
\end{figure}

\begin{figure}
    \centering\scalebox{0.95}{
    \includegraphics[width=0.45\textwidth]{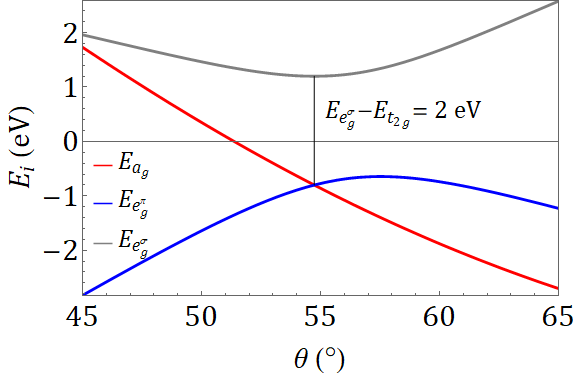}}
    \\
    \scalebox{0.95}{\includegraphics[width=0.45\textwidth]{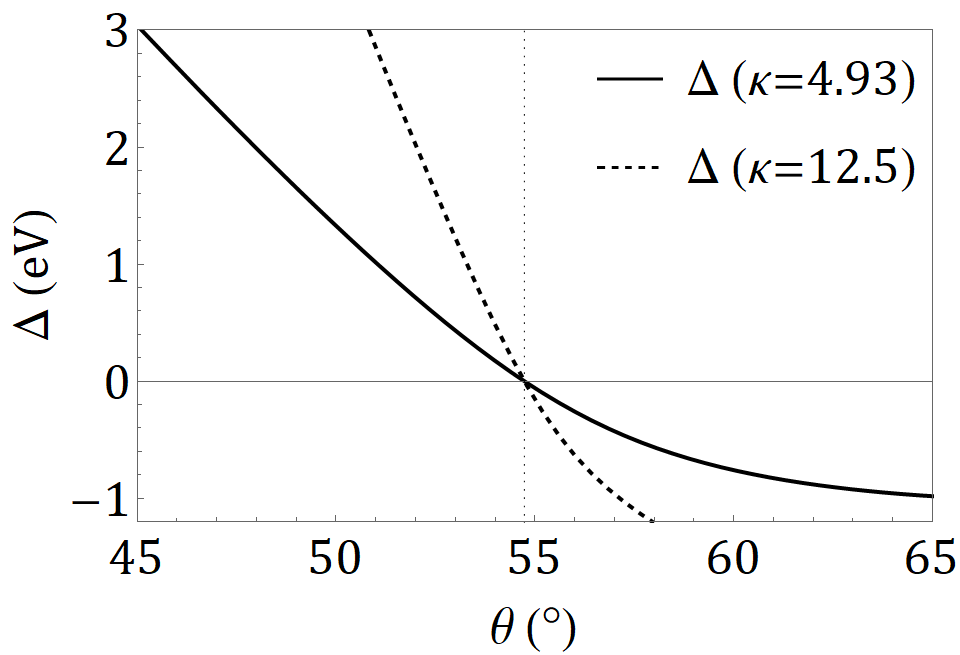}}
    \caption{(Upper panel) $a_g$, $e_g^{\pi}$ and $e_g^{\sigma}$ energy level behaviour as a function of the distortion angle $\theta$ using $E_0=2$ eV, $\kappa=4.93$ and $z^*=4$. (Lower panel) $\Delta$ parameter as a function of the distortion angle $\theta$ using the same parameters (solid line) and using $\kappa=12.5$, corresponding to the linearized muffin tin approximation (dashed lines).}
    \label{fig:comparison}
\end{figure}

We have shown that the possibility for the system to exhibit non trivial properties is critically connected with the strength of the strain parameter $\Delta$. One can therefore wonder whether it is possible under realistic experimental conditions to achieve the necessary values for the trigonal coupling in order to have $|\Delta/t_3|\gtrsim1$.
In Refs. \cite{khomskii2016role, kugel2015spin} an analysis of the trigonal strain is performed in the case of a face-sharing octahedral structure. In such a case a trigonal distortion is a compression or a stretching of the face-shared planes of the oxygen octahedra around a metal ion. When the material is unstrained the octahedral oxygen cage split up the $d$-orbitals into $t_{2g}$ and $e_g^\sigma$ orbitals. When a strain is imposed the angle M-O-M between two metal ions and a oxygen in between changes, reducing the symmetry of the system from octahedral to trigonal and splitting up the states into three subgroups: $e_{g}^\sigma$, $a_g$ and $e_g^\pi$ orbitals. The method of Refs. \cite{khomskii2016role, kugel2015spin} allows a unified treatment of the energy splitting caused by the trigonal distortion and recovers the energy splitting of the octahedral crystal field when the M-O-M angle takes its undistorted value.
The same calculation can be used for the case of a corner-sharing structure as in the STO case.
The strain at the interface is translated into a rigid shift of the atoms along the (111) direction. This leads to a distortion of the octahedral oxygen cage structure around one Ti atom, analogous to the case of the face-sharing octahedra. This leads to the energy splitting between the $a_g$, $e_g^{\pi}$ and $e_g^{\sigma}$ states depending on the distortion angle $\theta$ between the direction connecting the O ligand with the Ti ion line and the (111) direction as shown in Fig.~(\ref{fig:cage}).
This energy splitting explicitly depends on the following parameters: 
\begin{itemize}
    \item the energy splitting $E_0$ between the $e_g^{\sigma}$ states and the triplet of states $a_g$ and $e_g^{\pi}$, which are degenerate when $\Delta=0$. This splitting is typically expressed as $E_0=10Dq$ \cite{khomskii2014transition};
    \item $\kappa=a_0^2\frac{\langle r^2\rangle}{\langle r^4\rangle}$, where $a_0$ is the lattice constant and the mean values of the powers of the radii are evaluated between the orbitals of the conduction band;
    \item the effective nominal charge $z^*$ which affects the electrons in the conduction band. This is evaluated by subtracting from the ion charge $Z$ the charge of the electrons in the complete shells and of those electrons in the non-complete shells multiplied by a corrective factor ($0.35$), which takes into account the electron-electron screening.
\end{itemize}
Using these results for the STO structure, choosing $E_0=2$ eV as reported in Ref. \cite{mattheiss1972energy}, $\kappa=4.93$, as determined for the $d$ orbitals, and $z^*=22-18=4$, since Ti$^{4+}$ ions have no electrons in their external shells, we reproduce the energy dependence of the $a_g$, $e_g^{\pi}$ and $e_g^{\sigma}$ orbitals as a function of the distortion angle $\theta$, together with the behaviour of the strain parameter $\Delta=(E_{a_g}-E_{e_g^{\pi}})2/3$ in Fig.~(\ref{fig:comparison}). For the latter we show the results also for $\kappa=12.5$, corresponding to the muffin tin approximation \cite{kugel2015spin}.
The results show that for experimentally accessible distortion angles, $\Delta$ reaches absolute values sufficiently large to cause the appearance of the Dirac cone in the band structure. In fact, compared to the equilibrium position $\theta_0=\arccos{(1/\sqrt{3})}\approx54.7^{\circ}$ for which there is no splitting between the $a_g$ and $e_g^{\pi}$ states, for $\theta\geq58.7^{\circ}$ we find $\Delta\leq-0.6$ eV. The muffin tin approximation leads to even smaller values of distortion angle necessary to reach the same value of $\Delta$. For this reason for the rest of the paper, unless we specify otherwise, we consider as benchmark the value $\kappa=4.93$ which leads to conservative estimates for $\Delta$.
\\We observe that our results do not take into account the correction due to the crystal field originated by the metal ions in the structure; the introduction of such a correction only increases the strength of the trigonal coupling in absolute value. Our results are therefore meant as a lower bound on the realizable coupling $\Delta$, and they show that, already at this level, the trigonal dominated regime proposed in the previous sections seems to be experimentally achievable. These conclusions are further strengthened in Appendix \ref{Strainandhopping}, in which we show that they are robust by accounting for the modification of the hopping parameters induced by the distortion angle.
Let us finally discuss the change in the in-plane lattice constant induced by the biaxial strain along the (111) direction. 
As reported in Ref.~\cite{PhysRevMaterials.3.030601}, a biaxial (111) strain in the bulk STO could lead to transition to a ferroelectric phase if the strain is tensile. We first point out that, in the realistic case 
corresponding to $\Delta=-0.65$ eV and a distortion angle of 2$^\circ$ evaluated using the muffin tin approximation, 
the compression of the lattice constant is about $5\%$. In order to maintain the volume of the lattice cell constant, the in-plane hexagonal lattice has to expand by about $\sim2.5\%$ of the lattice constant along the $[\overline{1}10]$ and $[\overline{11}2]$ directions. This would result in a dilation of the side of the hexagon by the same relative amount, preserving the $C_{3v}$ symmetry of the 2D lattice. For this reason, we neglected this effect in our work. Furthermore, given that the considered strain for negative $\Delta$ is compressive, we expect a paraelectric phase for the STO, according to Ref.~\cite{PhysRevMaterials.3.030601}, which does not modify the predicted results, up to a redefinition of the electric field coupling $v$.
\section{Topological properties}\label{topological}
\begin{figure}[b!]
    \centering
    \includegraphics[width=0.35\textwidth]{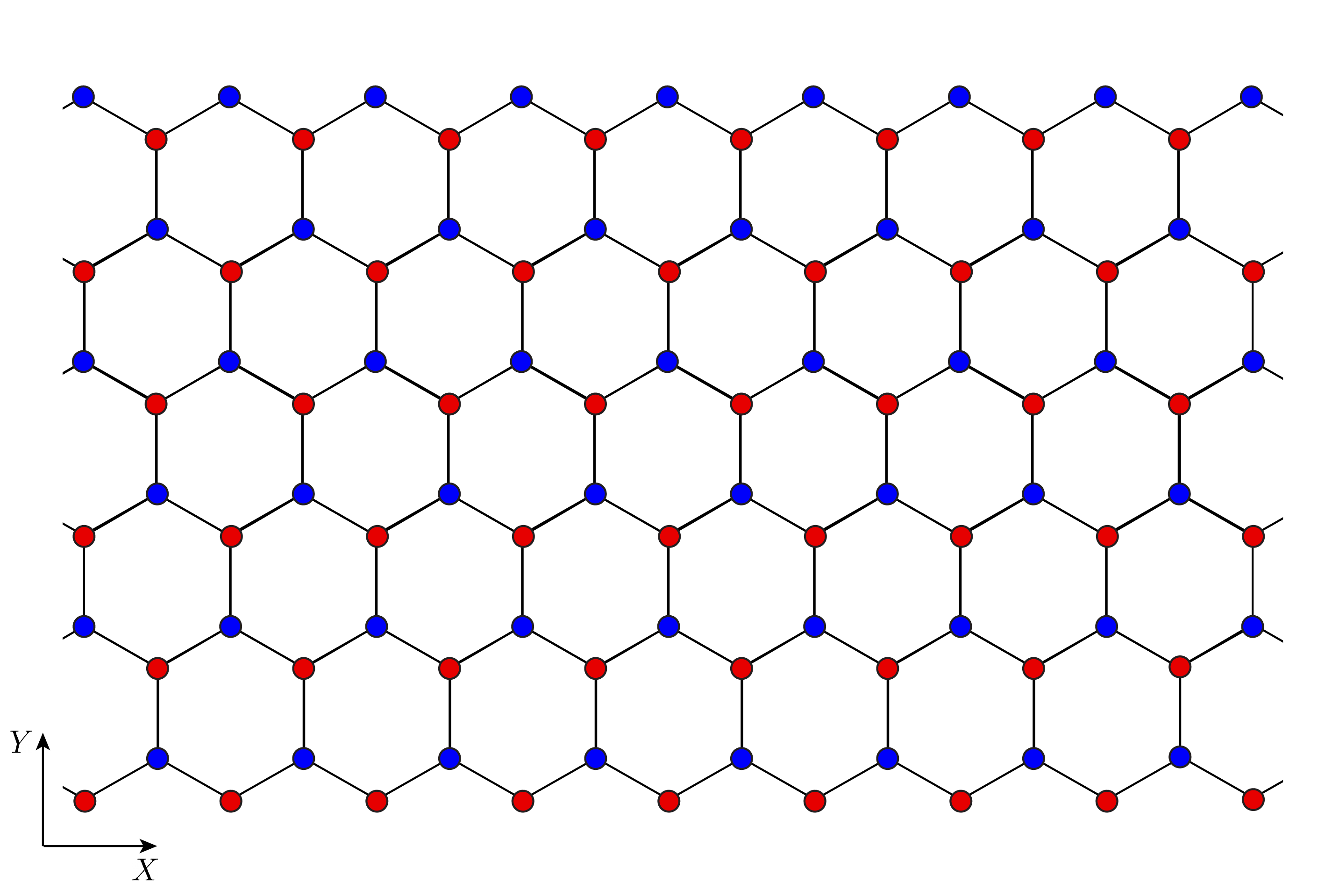}
    \caption{Schematic representation of a finite-thick zigzag ribbon. The system is confined along the vertical Y-axis while is unlimited along the X-axis.}
    \label{zigzagribb}
\end{figure}
\begin{figure}[t!]
    \raggedright
    \includegraphics[width=0.43\textwidth]{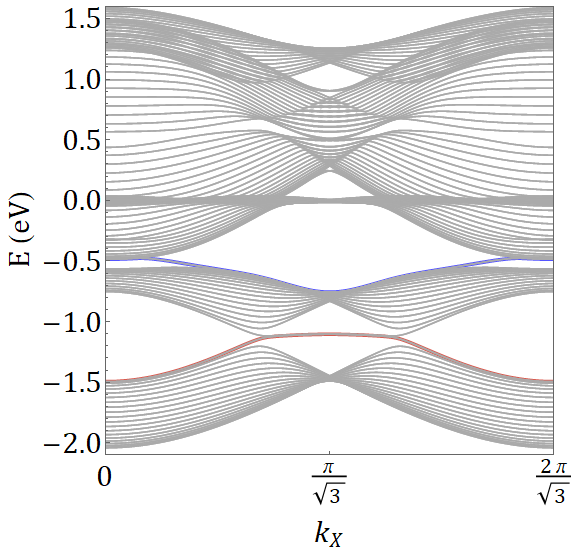}\hspace{0.5cm}
    \\
    \includegraphics[width=0.42\textwidth]{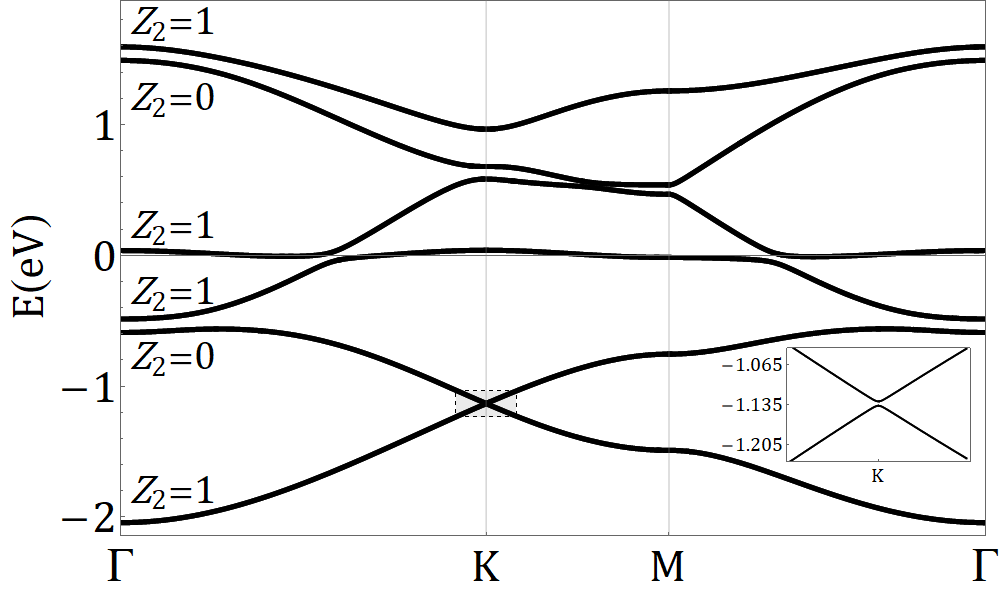}
    \\
    \includegraphics[width=0.43\textwidth]{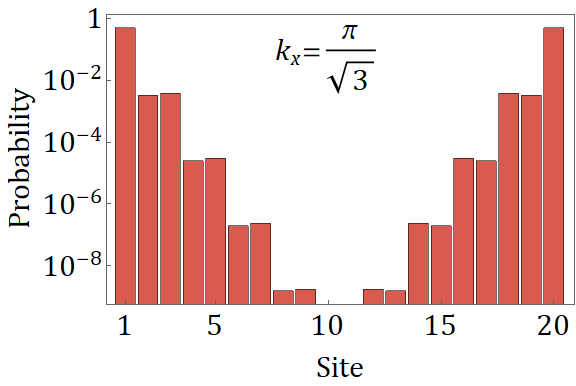}\hspace{0.5cm}
    \caption{(Upper panel) Energy dispersion for a finite-thick zigzag ribbon of 20 sites, for $\Delta=-1$ eV, $\lambda=0.1$ eV and $v=0$ eV. The colored lines represent the edge states of the system, where there is an energy band inversion within the gap. (Middle panel) Energy band structure for the same parameters and the corresponding $Z_2$ invariant for each band. (Lower panel) The probability of a carrier occupation of a site for $k_X=\pi/\sqrt{3}$ for the red curve in the upper panel.}
    \label{fig:nanoribbon}
\end{figure}
\begin{figure*}[t!]
    \centering
    \includegraphics[width=0.30\textwidth]{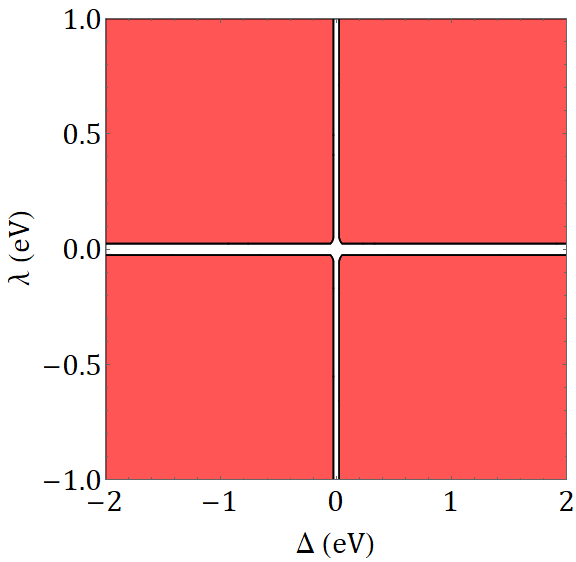}\hfill
    \includegraphics[width=0.30\textwidth]{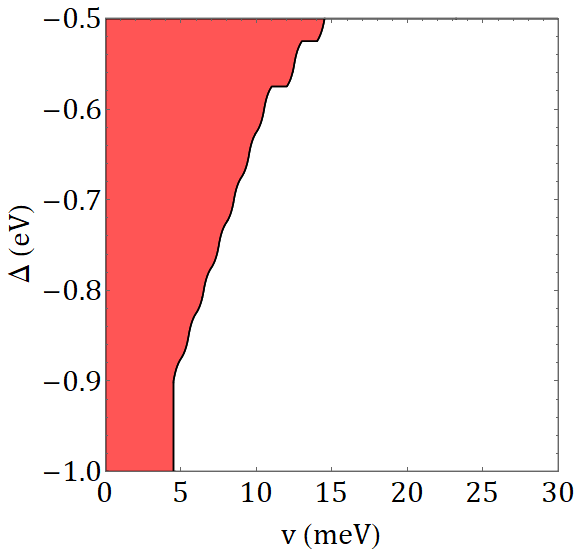}\hfill
    \includegraphics[width=0.30\textwidth]{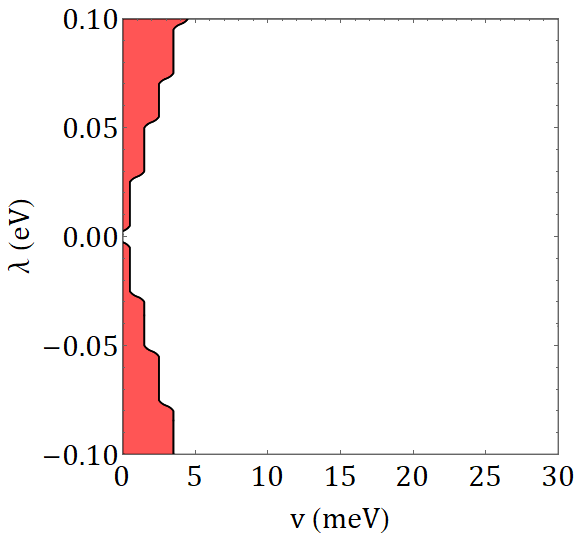}
    \caption{Phase diagram for the first Kramers doublet (a) in the $\lambda-\Delta$ plane with $v=0$ eV; (b) in the $v-\Delta$ plane fixing $\lambda=0.1$ eV and (c) in the $\lambda-v$ plane fixing $\Delta=-1$ eV. The red region corresponds to $Z_2=1$, while the white one corresponds to $Z_2=0$. The diagram (a) has been computed using an energy step of $0.05$ eV, the diagram (b) has been computed using an energy step for $\Delta$ ($v$) of $0.05$ ($0.001$) eV and the diagram (c) using an energy step for $\lambda$ ($v$) of $0.005$ ($0.001$) eV.}
    \label{contours}
\end{figure*}
Due to the simultaneous presence of the SOC, the Dirac cone in the electronic band structure and the time reversal invariance, we expect that, at least for $|\Delta/t_3|\gtrsim1$, the system could exhibit QSHE.
In this section we scan the parameter space of the system in order to determine which are the conditions for this to happen. 
\\Since in the strained interface all the conditions for QSHE are met, we simulated a finite-thick (20-100 sites) zigzag ribbon, whose representation is shown in Fig.~(\ref{zigzagribb}), obtaining numerical evidence of edge states.
The latter are represented by the band inversion in the energy dispersion along the $k_X$ coordinate, the momentum in the orthogonal direction to the finite length of the ribbon. The results for a 20 sites ribbon are represented in Fig.~(\ref{fig:nanoribbon}), where we highlight in color the edge states. Notice that the SOC is chosen to be $0.1$ eV for ease of visualization (the typical values are smaller \cite{monteiro2019band}).
We also show the value of the $Z_2$ invariant computed for each of the Kramers doublet according to the method of the Wannier Charge Centers (WCCs).
In this method we identify the fundamental direction of the reciprocal lattice, named $\lambda_1$ and $\lambda_2$, and we compute the WCCs, namely the mean value of the position $i\frac{\partial}{\partial \lambda_1}$, for both of the states of a Kramers doublet as a function of the periodic variable $\lambda_2$. For each doublet we subsequently count the number of discontinuities of the mean WCC along half a period of $\lambda_2$. The sum modulo two of the number of discontinuities is equal to $Z_2$ invariant. The specific details of how to choose a continuous gauge along the BZ are discussed in more details in Refs. \cite{soluyanov2011wannier,soluyanov2011computing}.
The edge states are always formed in between the non trivial bands. The localized nature of these states is confirmed by the rapid decreasing of the occupancy probability away from the edge for the red band. The choice of showing the results for a 20 sites ribbon is to highlight this rapid decreasing. For a 100 sites ribbon we find, as expected, the same qualitative behavior. A particularly interesting conclusion is the fact that edge states can form at the K point, in between the Dirac cone; this constitutes a further similarity between our heterostructure and graphene.
\\The general topological characterization of the system has been obtained by computing the $Z_2$ topological invariant for each of the six Kramers doublets and subsequently determining the phase diagrams in the $\lambda-\Delta$ plane. In the left panel of Fig.~(\ref{contours}) we show the phase diagram for the lowest energy doublet. The diagrams for the other doublets can be found in the Appendix \ref{AppendiceFigure}. We can deduce from the figure that the first energy doublet is always non-trivial when $\lambda$ and $\Delta$ are non-zero.
\\In order to achieve a QSHE the necessary requirements a chemical potential within the energy gap in the band structure and a value equal to 1 for the $Z_2$ invariant; the latter is the sum modulo 2 of the $Z_2$ of the filled doublet. A topologically non-trivial behaviour is therefore expected when the chemical potential lies between the first and second Kramers doublet. We emphasize the behaviour of the first doublet, since it is easier to access experimentally, but, referring to Fig.~(\ref{fig:nanoribbon}), a non trivial behaviour is expected also if the bands are filled up to the second doublet.
\\Finally we studied the stability of the topological properties against an external electric field. In order to do this we evaluated the $Z_2$ invariant for a fixed value of $\lambda=0.1$ eV, varying $v$ and the trigonal parameter $\Delta$ in a region in which the system exhibits an insulating behaviour. 
The results are depicted in middle panel of Fig.~(\ref{contours}), showing how the topological structure for the lowest bands is preserved for electric potentials of $v\sim0.004$ eV for $\lambda=0.1$ eV. This corresponds to an electric field between the two layers of Ti of the order of $10^7$ V/m over a thickness of $\tilde a\sim0.3$ nm. Generally with the decreasing of the strain $|\Delta|$ a larger field is needed to bring the system into the normal state. We studied the stability against weak perturbations also varying the spin-orbit coupling, since it could in principle be different from the value used in the previous analysis: we fixed the trigonal parameter to $\Delta=-1$ eV obtaining the phase diagram shown in the right panel of Fig.~(\ref{contours}).
\begin{figure}
    \centering
    \includegraphics[width=0.45\textwidth]{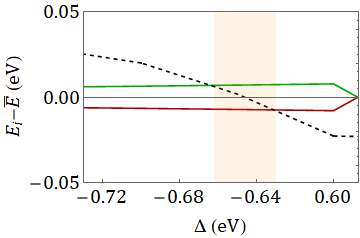}
    \caption{Energy gap between the first and the second Kramers doublet as a function of the strain parameter $\Delta$. The green (red) curve represents the minimum (maximum) energy $E^{II}_{min}$ ($E^I_{Max})$ of the upper (lower) Kramers doublet after subtraction of the quantity $\overline{E}=(E^I_{Max}+E^{II}_{min})/2$; the dashed line is the chemical potential $\mu-\overline{E}$ with a fixed surface density.
    The orange section is the region in which the system exhibits QSHE.}
    \label{fig:topotrans}
\end{figure}
In order to prove the possibility to induce a topological transition \textit{via} modification of the strain we show in Fig.~(\ref{fig:topotrans}) the energy gap between the bands as a function of the strain parameter, together with the chemical potential for a fixed value of the surface carrier density $n_{2D}$. We have chosen this density value so that the chemical potential $\mu$ lies within the energy gap for $\Delta=-0.65$ eV and for benchmark values of $\lambda=0.1$ eV and $v=0$ eV. From Fig.~(\ref{fig:topotrans}) it is evident that decreasing $\Delta$, and therefore increasing the strain, the chemical potential moves through the bands passing in the orange region. Here the two conditions of a non-trivial topological structure for the lower band and an insulating regime for the material are simultaneously met, so that the system can exhibit a QSHE.  
\section{Conclusions and discussion}\label{conclusioni}
In this work we investigated the properties of the (111) STO-based heterostructure interface via TB methodologies. 
We took into account the contribution of the electron hopping in the bilayer of STO at the surface, the interfacial trigonal strain, the SOC and the electric field orthogonal to the bilayer. The values of the main hopping parameters were obtained by comparison with the ARPES experiments. The remaining parameters were systematically studied in order to identify the features of the electronic structure. We identified two different regimes: one in which the strain is a small perturbation compared to the hopping parameters and the other in which it is a relevant contribution. In the former situation our model is able to capture the main features of the band structure previously obtained in literature \cite{khanna2019symmetry,rodel2014orientational,walker2014control}. In the latter the system exhibits a Dirac point in the structure of the low energy bands, for experimentally accessible values of the carrier density. Compared with Ref. \cite{doennig2013massive}, which obtained the same behaviour with DFT methodologies, we accurately discussed the role of SOC and assessed the regime of the trigonal strain where a Dirac cone is expected. We found that the presence of the Dirac point requires values of the trigonal distortion such that $|\Delta/t_3|\ge1$. Our assumption was supported by a systematical study of the dependence of the energetic parameter $\Delta$ on the physical strain applied at the interface. We also showed the presence of non trivial spin and orbital angular momentum patterns in the BZ, finding that in the presence of the trigonal distortion there can be a non vanishing magnetic dipole moment between the two layers of Ti at a fixed $\vek$ in the Brillouin zone. In agreement with the expectation of time reversal invariance this magnetic moment vanishes when integrated over the whole Brillouin Zone; nevertheless these results are potentially accessible using spin-polarized ARPES experiments. Interestingly in the regime with sizable strain, when the chemical potential is within the gap at the Dirac point, the out-of plane components of the spin are completely localized near the $K$ point in the Brillouin zone. These features suggest a non-trivial spin behaviour and open up the way to a systematic analysis of the spin properties of (111) STO heterostructures. We determined these properties in a minimal model using only two layers of Ti atoms. The presence of thin films over the STO, in particular LAO layers, are included as a \textit{reservoir} of electrons, and its possible role in generating distortion is completely contained within the strain parameter $\Delta$ and $v$.
Moreover, we did not include in our study the Coulomb interaction. This allowed to obtain flexible results from the model with small computational effort: indeed we plan to use this model as a starting point for studying different phenomena. Furthermore, we expect the Coulomb interaction to be weak in the low filling region. Its effect should mainly be a renormalization of the energy spectra for the occupied bands~\cite{khanna2019symmetry}, increasing locally the Fermi velocity. This conclusion should still be valid in the regime of interest $|\Delta|/t_3>1$ near the Dirac point, as shown in Ref.~\cite{sinner2010effect}. Here it is shown that the Coulomb interaction cannot open a gap in the band structure by itself, even though it widens the energy gap opened by SOC. Therefore, the main effects of Coulomb interactions that are not included in our work should be a renormalization of the Fermi velocity and the energy gap induced by SOC at the Dirac point. 
We point out that we did not investigate a poissible anti-ferrodistortive transition which can occur at low temperature in the bulk STO, and using (111) biaxial strain as predicted in Ref.~\cite{PhysRevMaterials.3.030601}. Moreover we did not investigate the possible presence of charge ordered modes as predicted in \cite{doennig2013massive} under suitable conditions. These questions can be investigated in future studies.

The low energy region in the regime with sizable strain is particularly interesting also due to its non-trivial topological properties. We computed the topological invariant of the system using the WCC method proposed by Soluyanov and Vanderbilt \cite{soluyanov2011wannier,soluyanov2011computing}. We found that the first Kramers doublet is characterized by a $Z_2=1$ in the presence of the inversion symmetry. This property is preserved in the presence of weak perturbations due to an external electric field. Finally, we showed that the system can exhibit a QSHE; this property has been investigated via finite-thick zigzag ribbon simulations, which show that edge states can be present under suitable conditions at the Dirac point. We found that a topological transition can occur since the system can be tuned from a conducting to an insulating state \textit{via} strain manipulation.

In conclusion, as a result of a sizable strain,  our  TB scheme predicts a  non-trivial topological phase for LAO/STO (111) including a gap at the Dirac point. The possibility to engineer the applied trigonal strain at the interface suggests that the transition to the non-trivial topological phase can be controlled in a parameter range which is experimentally achievable.

\section*{Acknowledgments}
C.A.P. acknowledges interesting discussions with A. Caviglia and G. De Luca.

This work was supported by the project QUANTOX (QUANtum Technologies with 2D-OXides) of QuantERA-NET Cofund in Quantum Technologies, implemented within the EU-H2020 Programme, and the project TOPSPIN (Two-dimensional Oxides Platform for SPINorbitronics nanotechnology) funded by the MIUR-PRIN Bando 2017 - grant 20177SL7HC.
\appendix
\section{Maximum likelihood analysis for $t_1$, $t_2$, $t_3$ parameters}\label{appendiceLikelihood}
\begin{figure}[t!]
    \centering
    \includegraphics[width=0.45\textwidth]{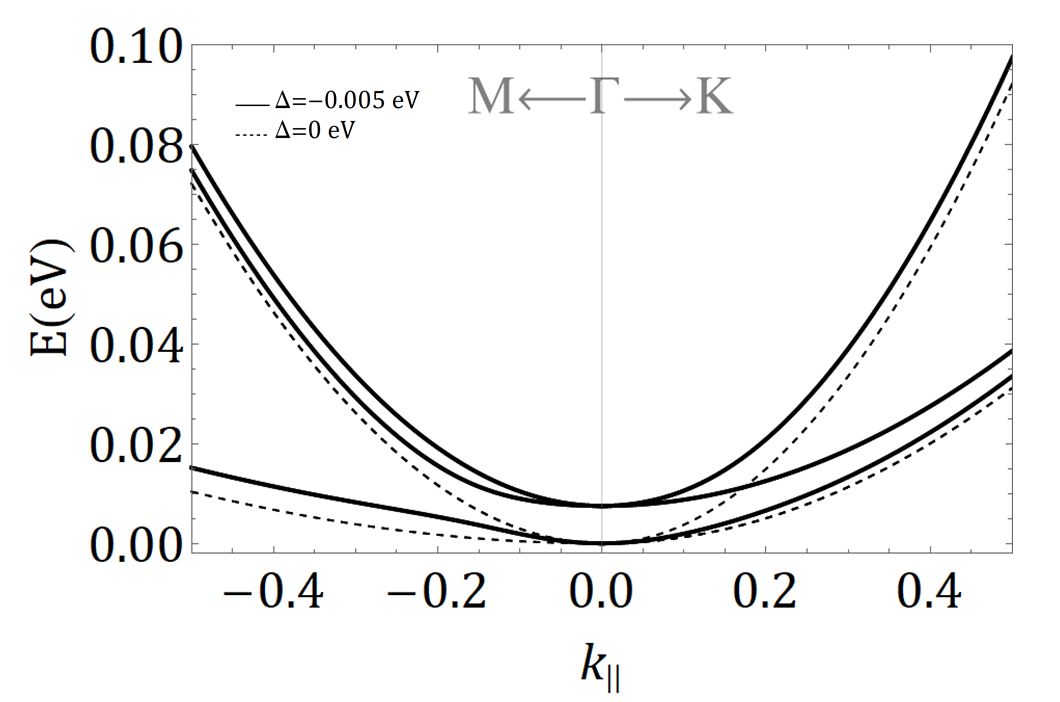}
    \caption{Detail of the bottom of the bands near the $\Gamma$ point, along the line connecting $M$ with $\Gamma$ ($k_Y$ direction) on the left side and the line connecting $\Gamma$ with $K$ ($k_X$ direction) on the right side. The dashed line represents the bands using only the TB model while the solid line includes a small trigonal strain parameter: the splitting induced by the latter is of the order of $3/2 \Delta$. The minima of the bands in the two cases have been shifted to coincide for ease of visualization.}
    \label{Arpes}
\end{figure}
\begin{figure}[t]
    \centering
    \includegraphics[width=0.45\textwidth]{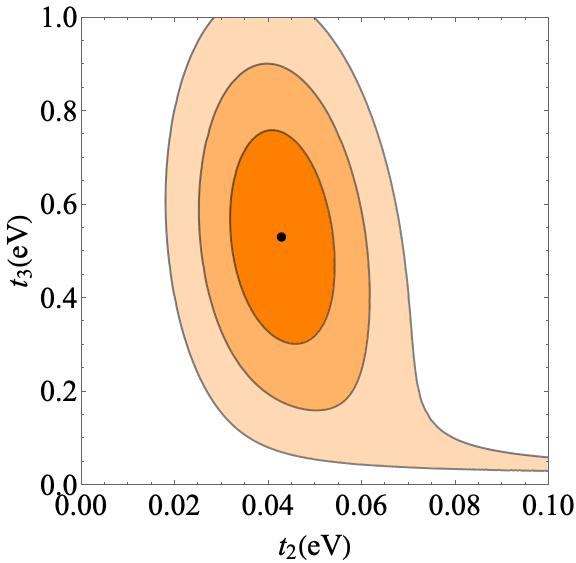}
    \caption{Contours of exclusion at $68\%$, $95\%$ and $99.7\%$ confidence level in the $t_2$-$t_3$ plane: the best fit point is marked in black.}
    \label{chi}
\end{figure}
In this appendix we discuss the methodology used to extract the values of the hopping parameters of the matrix $t_i^{\alpha\beta}$ in Eq.~(\ref{TBk}).
In literature STO bare surface ARPES data are available \cite{walker2014control} which allow us to fix the hopping parameters $t_1$, $t_2$ and $t_3$, since the qualitative behaviour of the interface depends on the ratio between them and $\Delta$. As a first approximation we can fix $\Delta=0$, since its value is of the order of the meV for the bare STO \cite{de2018symmetry}, $\lambda=0$ and $v=0$. In this framework we can extrapolate analytically the eigenvalues of the system Hamiltonian, reported here in a matrix form for simplicity\footnote{The double spin degeneracy is understood.}:
\begin{equation}
    \begin{aligned}
H&=\sum_{\vek}\\
    &\begin{pmatrix}
    d_{yz\text{Ti}_1,\vek}^\dagger\\
    d_{zx\text{Ti}_1,\vek}^\dagger\\
    d_{xy\text{Ti}_1,\vek}^\dagger\\
    d_{yz\text{Ti}_2,\vek}^\dagger\\
    d_{zx\text{Ti}_2,\vek}^\dagger\\
    d_{xy\text{Ti}_2,\vek}^\dagger\\
    \end{pmatrix}^T
    \begin{pmatrix}
        \gamma_{yz} & 0 & 0 & \epsilon_{yz} & 0 & 0 \\
        0 & \gamma_{zx} & 0 & 0 & \epsilon_{zx} & 0 \\
        0 & 0 &\gamma_{xy} & 0 & 0 & \epsilon_{xy} \\
        \epsilon_{yz}^* & 0 & 0 & \gamma_{yz} & 0 & 0 \\
        0 & \epsilon_{zx}^* & 0 & 0 & \gamma_{zx} & 0 \\
        0 & 0 & \epsilon_{xy}^* & 0 & 0 & \gamma_{xy} \\
    \end{pmatrix}
    \begin{pmatrix}
    d_{yz\text{Ti}_1,\vek}\\
    d_{zx\text{Ti}_1,\vek}\\
    d_{xy\text{Ti}_1,\vek}\\
    d_{yz\text{Ti}_2,\vek}\\
    d_{zx\text{Ti}_2,\vek}\\
    d_{xy\text{Ti}_2,\vek}\\
    \end{pmatrix}
    \label{TBmatrix},
    \end{aligned}
\end{equation}
where the intralayer, diagonal, contributions are:
\begin{equation}
    \begin{aligned}
    &\gamma_{yz}=-2t_1\cos(\frac{-\sqrt{3}}{2}k_X+\frac{3}{2}k_Y),
    \\
    &\gamma_{zx}=-2t_1\cos(\frac{\sqrt{3}}{2}k_X+\frac{3}{2}k_Y),
    \\
    &\gamma_{xy}=-2t_1\cos(\sqrt{3}k_X),
    \label{intralayer}
\end{aligned}
\end{equation}
and the interlayer contributions are:
\begin{eqnarray}
     \nonumber &\epsilon_{yz}&=-t_3\left(1+e^{i(\frac{\sqrt{3}}{2}k_X-\frac{3}{2}k_Y)}\right)-t_2e^{-i(\frac{\sqrt{3}}{2}k_X+\frac{3}{2}k_Y)},\\
     &\epsilon_{zx}&=-t_3\left(1+e^{-i(\frac{\sqrt{3}}{2}k_X+\frac{3}{2}k_Y)}\right)\\ \nonumber &-&t_2e^{i(\frac{\sqrt{3}}{2}k_X-\frac{3}{2}k_Y)},\\
     \nonumber &\epsilon_{xy}&=-2t_3\cos(\frac{\sqrt{3}}{2}k_X)e^{-i\frac{3}{2}k_Y}-t_2,
     \label{interlayer}
\end{eqnarray}
where $\vek=k_X \hat{u}_{\overline{1}10}+k_Y\hat{u}_{\overline{1}\overline{1}2}$ is the dimensionless in-plane $k$ vector of the Brillouin Zone. We emphasize that the coupling among different $t_{2g}$ orbitals are all equal by virtue of the $C_3$ symmetry of the interface.
The possibility of exact diagonalization is evident due to the fact that the tight binding interaction does not mix the different $d_i$ orbitals with one another, but only the two layers. The matrix is therefore in a $2\times2$ block diagonal form. 
The eigenvalues, which are doubly degenerate in the spin, are the following:
\begin{equation}
    \rho_i(\Vec{k})=\gamma_i(\Vec{k})\pm|\epsilon_i(\Vec{k})|
    \label{eigenTB}
\end{equation}
where $i$ runs over $yz,zx,xy$, $\gamma_i(\Vec{k})$ in which we have explicitly expressed the $\Vec{k}$ dependence.
In $\Vec{k}=0$ ($\Gamma$ point in the Brillouin Zone) the eigenvalues referring to the lower band ($-$ sign) have a threefold degenerate minimum (neglecting the spin degeneracy)
\begin{equation}
    E_ {min}=2t_1-(2t_3+t_2).
\end{equation}
Near the bottom of the bands an approximate expression for the energies is
\begin{equation}
    \rho_i(k_X,k_Y)=E_{min}+\frac{\hbar^2 k_X^2}{2(m^*_{\Gamma K})_i}+\frac{\hbar^2 k_Y^2}{2(m^*_{\Gamma M})_i}+\alpha_i k_X k_Y,
    \label{roh}
\end{equation}
where we defined
\begin{eqnarray}\label{effmasses}
     \nonumber &(m^*_{\Gamma K})_{yz}=(m^*_{\Gamma K})_{zx}=\frac{4\hbar^2}{3 (2 t_1 + \frac{t_3 (5 t_2 + t_3)}{t_2 + 2 t_3})};\\
     &(m^*_{\Gamma M})_{yz}=(m^*_{\Gamma M})_{zx}=\frac{4\hbar^2}{9 (2 t_1 + \frac{t_3 (t_2 + t_3)}{t_2 + 2 t_3})};\\
     \nonumber &(m^*_{\Gamma K})_{xy}=\frac{2\hbar^2}{3 (4 t_1 + t_3)}; \quad (m^*_{\Gamma M})_{xy}=\frac{2\hbar^2 (t_2+2t_3)}{9 t_2t_3}.
\end{eqnarray}
These quantities represent the effective masses in the free particle approximation along the directions connecting respectively the $\Gamma$ and $K$ points and the $\Gamma$ and $M$ points in the Brillouin zone.
The numerical values of the parameters have been fixed using the ARPES data \cite{walker2014control} shown in Fig.~(\ref{Arpes}): from the dashed curves we have extrapolated the values of the effective masses, together with a confidence range. In this way, a $\chi^2$ analysis has been performed in order to search for the best fit combination of the three parameters.
\begin{figure*}
\centering
    \includegraphics[width=0.8\textwidth]{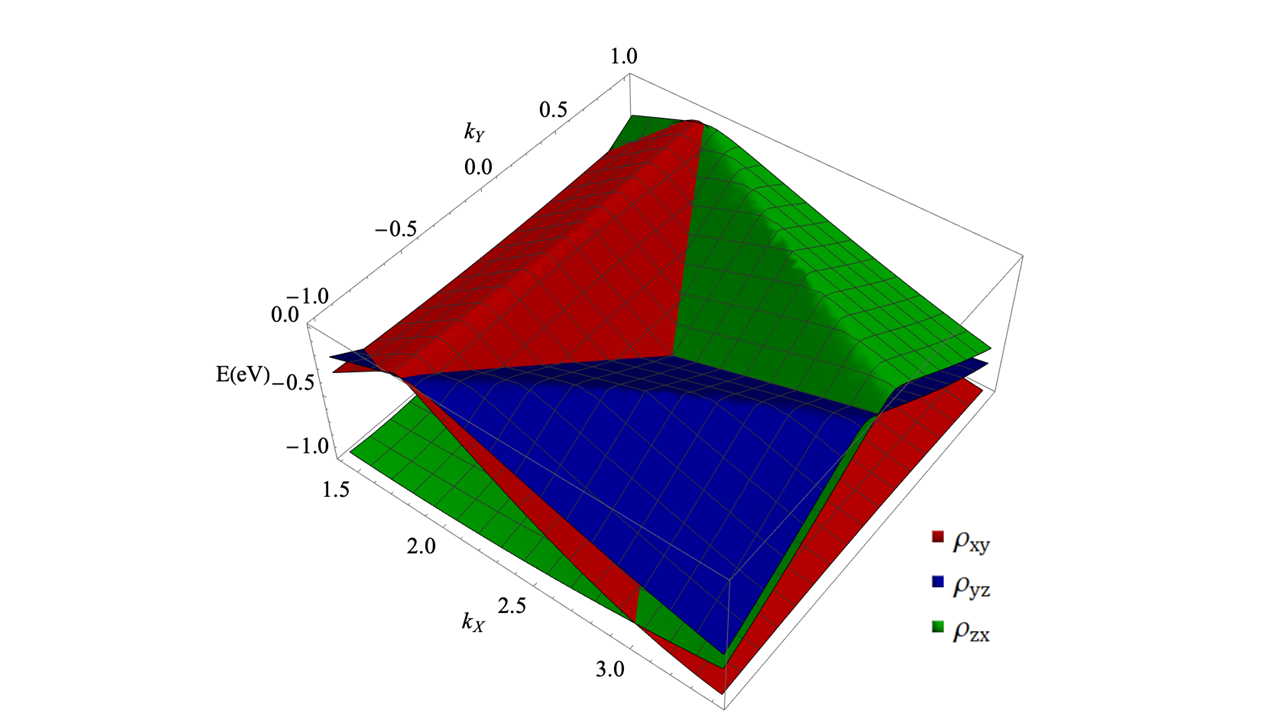}
    \caption{Structure of the lower bands near their intersection at the $K$ point: the bands behave linearly as planes near the $K$ point.}
    \label{Kplotanalytic}
\end{figure*}
The best fit values are:
    \begin{equation}
        t_3=0.5\hspace{0.2cm}{\rm{eV}}; 
        \quad t_2=0.04\hspace{0.2cm} {\rm{eV}}; 
        \quad t_1=0\hspace{0.2cm} {\rm{eV}}.
    \end{equation}
We do not provide the confidence interval since this is meant only to be a qualitative analysis. Nonetheless we show in Fig.~(\ref{chi}) the $\chi^2$ contours in the $t_3$-$t_2$ plane. From the contours we see that the number of digits for the parameters is chosen coherently with their uncertainty on them.
Analogous studies in literature \cite{rodel2014orientational} have found similar values for the parameters.
\\In Fig.~(\ref{Arpes}) (solid line) we show the effect of a weak trigonal strain on the pure TB curves. The strain changes the effective masses near the $\Gamma$ point and induces a splitting between the first and the other two doublets of the order of $3/2\Delta$. However, the splitting is not sufficient to induce an effective gap between the first two doublets and the third one.
\section{Effective model near the $K$ point for the unstrained interface} 
\begin{figure}
    \centering
    \includegraphics[width=0.45\textwidth]{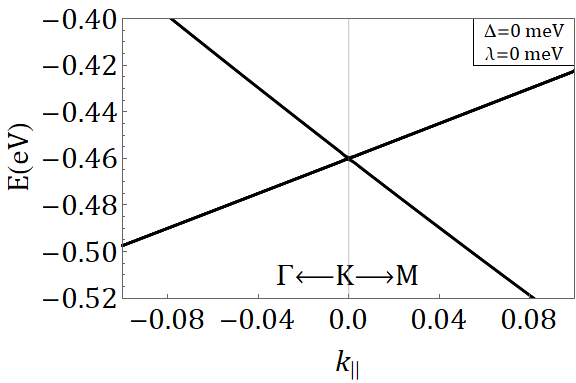}
    \hfill
    \includegraphics[width=0.45\textwidth]{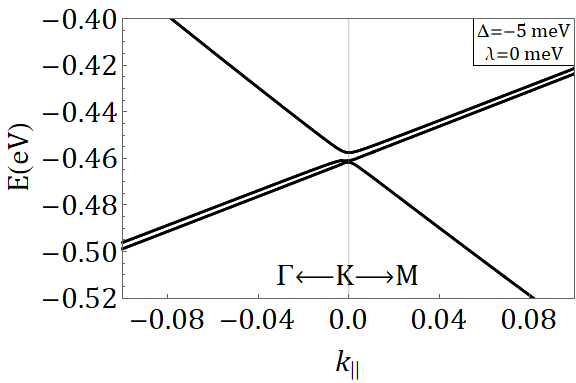}
    \caption{(Upper panel) Detail of the band structure near the $K$ point for the pure tight binding model. (Lower panel) Same as above in the presence of a trigonal crystal field $\Delta=-0.005$ eV: the degeneracy is partially split.}
    \label{DeltaPlotK}
\end{figure}

\begin{figure}
    \centering
    \includegraphics[width=0.45\textwidth]{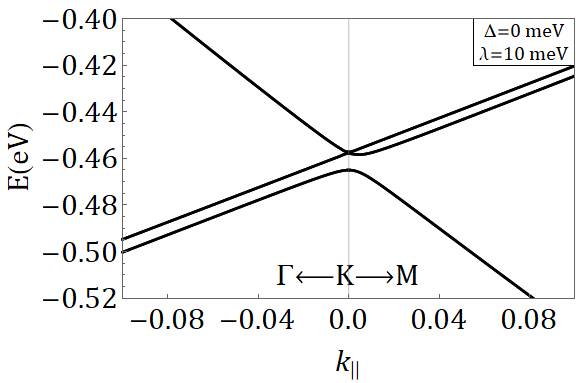}
    \hfill
    \includegraphics[width=0.45\textwidth]{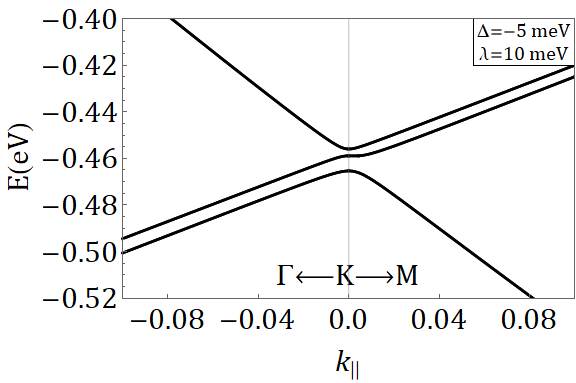}
    \caption{(Upper panel) Detail of the band structure near the $K$ point in the presence of a spin-orbit coupling $\lambda=0.01$ eV: the degeneracy is partially split. (Lower panel) Same as before, with the simultaneous presence of $\Delta=-0.005$ eV and $\lambda=0.01$ eV.}
    \label{lambdaKplot}
\end{figure}

In the case of unstrained interface, the value of $\Delta$ is fixed to be $-0.005$ eV. As we discussed above, at the $K$ point there is an intersection of bands that, along the high symmetry lines, has a structure similar to the Dirac point in the graphene. We will now detail the differences between our case and the graphene one.
\\In the same way as we did for the $\Gamma$ point in Appendix \ref{appendiceLikelihood}, we will provide an analytical approximation for the energy eigenvalues in the case of pure TB. 
\\The 3D structure nearby $K$, corresponding to $\Vec{k}=(\frac{4\pi}{3\sqrt{3}},0)$, is represented in Fig.~(\ref{Kplotanalytic}) for the lower bands where we set $\Delta=0$. The structure represents the intersection of three planes in the $K$ point, which is different from the Dirac cone.
Thus nearby the $K$ point the eigenvalues in Eq.~(\ref{eigenTB}) can be expanded both for the upper and the lower bands to first order as
\begin{equation}
    \rho_i \simeq \rho_{0}+\hbar \Vec{v}_{Fi}\cdot (\Vec{k}-\Vec{k}_0),
\end{equation}
where $\Vec{k}_0=(\frac{4\pi}{3\sqrt{3}},0)$, and $\rho_0=t_1\pm|t_2-t_3|$ with the minus (plus) sign referring to the lower (upper) bands.
We will focus on the lower bands; the values of $\Vec{v}_{Fi}$ depend on the orbital under examination:
    \begin{eqnarray}
        \nonumber &\Vec{v}&_{Fyz}=\\ \nonumber &\frac{1}{\hbar}&\left(\frac{3}{4}\left(2t_1+\frac{t_3(t_3- t_2)}{\sqrt{(t_3-t_2)^2}}\right),\frac{3\sqrt{3}}{4}\left(2t_1+\frac{t_3\sqrt{(t_3-t_2)^2}}{t_2-t_3}\right)\right);\\ 
        &\Vec{v}&_{Fzx}=\\ \nonumber &\frac{1}{\hbar}&\left(\frac{3}{4}\left(2t_1+\frac{t_3(t_3-t_2)}{\sqrt{(t_3-t_2)^2}}\right),\frac{3\sqrt{3}}{4}\left(2t_1+\frac{t_3(-t_2+t_3)}{\sqrt{(t_3-t_2)^2}}\right)\right); \\ \nonumber
        &\Vec{v}&_{Fxy}=\frac{1}{\hbar}\left(-3t_1+\frac{3}{2}\frac{t_3(t_2- t_3)}{\sqrt{(t_3-t_2)^2}},0\right).
    \end{eqnarray}
It is possible to write an effective Hamiltonian in the proximity of the $K$ point. The eigenvectors of the $2\times2$ matrices of Eq.~(\ref{TBmatrix}) are
\begin{equation}
    \ket{\psi_{i,k}^{\pm}}=\frac{1}{\sqrt{2}}\left(\ket{\psi_{i\text{Ti}_1,k}}\pm e^{i\varphi_i(\Vec{k})}\ket{\psi{_{i\text{Ti}_2,k}}}\right)
    \label{eigenK},
\end{equation}
with the sign referring to the lower or the upper bands, $e^{i\varphi_i(\Vec{k})}=\frac{|\epsilon_i(\Vec{k})|}{\epsilon_i(\Vec{k})}$. We have neglected the spin index for simplicity.
Thus we can write the effective Hamiltonian as
\begin{equation}
    H_K=\sum_{\Vec{k}\simeq\Vec{k}_0}\sum_{i,\lambda,\sigma}\lambda\hbar\Vec{v}_{Fi}^\lambda \cdot(\Vec{k}-\Vec{k}_0)d_{i\sigma,k}^{\lambda\dagger}d_{i\sigma,k}^{\lambda},
\end{equation}
with $\lambda=\pm$ discriminating between the upper or the lower bands, and $d_{i,k}^{\lambda\sigma}$ is the annihilation operator referring to the states in Eqs.~(\ref{eigenK}).
\\We conclude that the $K$ point in this case is quite different from the Dirac point in graphene: in the latter, in fact, there are no states near the Dirac point, while in our case there is a large quantity of states corresponding to the intersection of the three plane bands. This is also seen from the density of states in Fig.~(\ref{fig:BandDSmallNoLNov}) at the dashed line. In order to experimentally realize an electron filling up to the $K$ point an electron surface density of $n_{2D}=1.59\times10^{15}$ cm$^{-2}$ is needed, which is large compared with the typical charge carriers densities measured or predicted at the interface under examination.
\\We will now discuss the stability of the degeneracy of the $K$ point upon including the perturbations. First of all, using the trigonal crystal field coupling of $\Delta=-0.005$ eV the band structure changes as in Fig.~(\ref{DeltaPlotK}).
The degeneracy is partially split, without the formation of a gap.
\\The presence of $v$ does not influence the existence of the degeneracy: its effect is limited to pushing the upper and lower band further and further apart.
\\The spin-orbit coupling, on its own, is not able to remove the degeneracy as shown in Fig.~(\ref{lambdaKplot}). On the other hand the simultaneous presence of the trigonal field and the spin orbit interaction removes completely the degeneracy which leads to the intersection of the bands. Of course, for sufficiently small values of $\lambda$ and $\Delta$, the effect is limited only to an extremely narrow region near $K$, and the behaviour returns to be quasi-linear as we go far from the $K$ point.
\label{unstrained section}
\section{Effective Dirac Hamiltonian near the $K$ point for strained interface}\label{appendice LARGE DELTA}
\begin{figure*}[t!]
    \centering
    \includegraphics[width=0.23\textwidth]{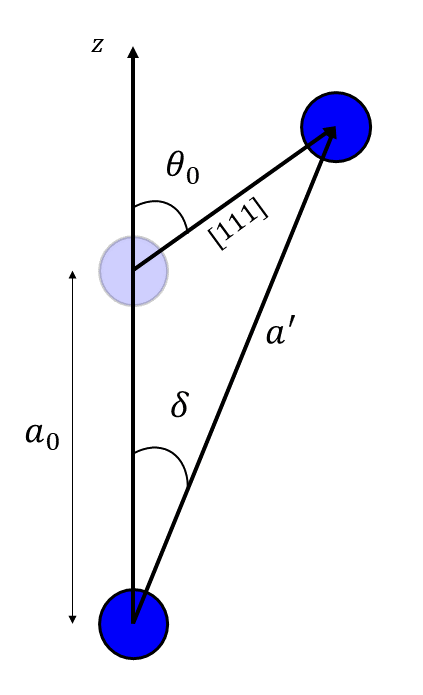}
    \hspace{2.8cm}
    \includegraphics[width=0.35\textwidth]{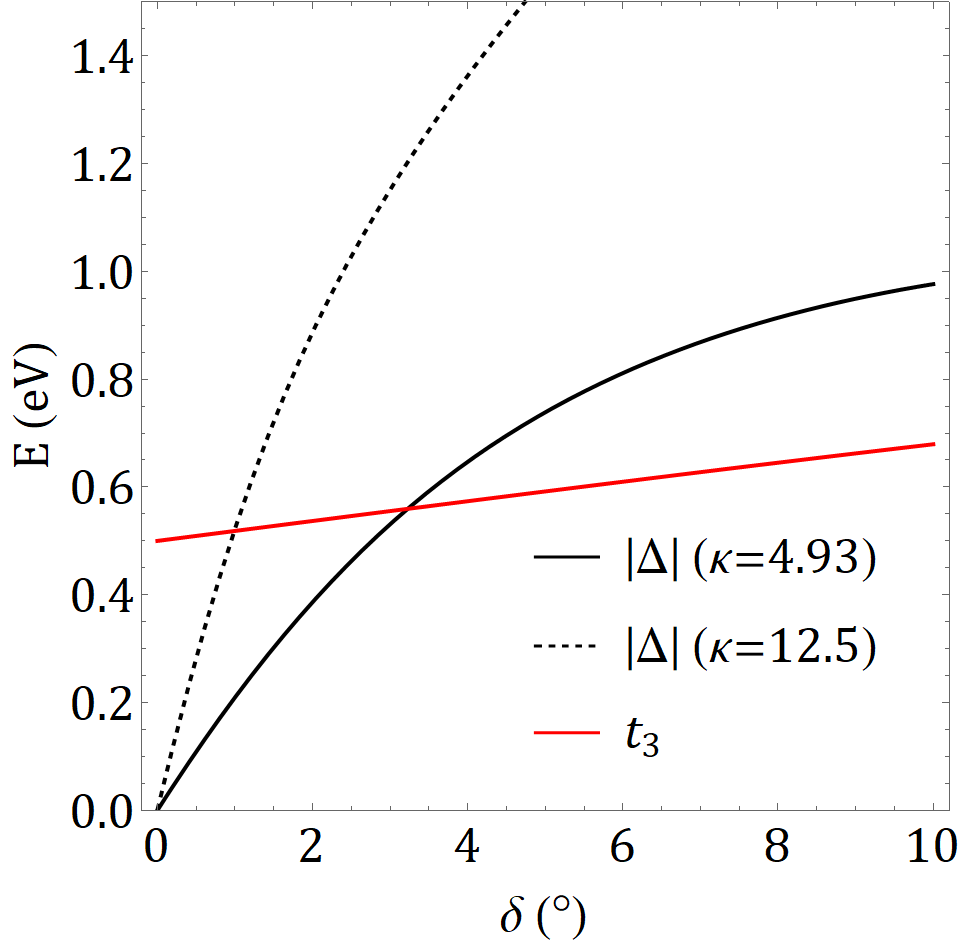}
    \caption{(Left panel) Geometrical representation of the strain $\delta$ and the corresponding change in the bond length between two Ti: the angle $\theta_0=\arccos(1/\sqrt{3})\sim54.7^\circ$. (Right panel) Dependence of the TB parameters on the distortion angle $\delta$. We show the behavior both of the strain parameter $|\Delta|$, for two benchmark values of $\kappa$ (black, solid and dashed), and of the hopping parameter $t_3$ (red) as a function of $\delta$.}
    \label{angolo_strain}
\end{figure*}
The case of the strained surface is particularly interesting due to the presence of the structure resembling the Dirac Point. Since $\Delta=-1$ eV it is possible to analyze the proximity of the point thanks to a perturbative approach in which the tight binding terms are treated as small. In fact we can consider the two lower bands as originating only from the $\ket{a_g\text{Ti}_1,\vek}$ and $\ket{a_g\text{Ti}_2,\vek}$ Bloch orbitals belonging to layer Ti$_1$ and Ti$_2$: this would be an exact statement in the case of infinite $\Delta$. Even though $\Delta=-1$ eV is not exactly in the range of applicability of perturbation theory (because $t_3=0.5$ eV), this would be enough to give us a semi-quantitative understanding of what happens at the K point.
\\In the subspace of degeneracy constituted by the two states $\ket{a_g\text{Ti}_1,\vek}$ and $\ket{a_g\text{Ti}_2,\vek}$ the perturbation induced by the TB is
\begin{equation}
    H_{{\rm{eff}}}=\frac{1}{3}\begin{pmatrix}
    \gamma_{yz}+\gamma_{zx}+\gamma_{xy} & \epsilon_{yz}+\epsilon_{zx}+\epsilon_{xy}\\
    \epsilon_{yz}+\epsilon_{zx}+\epsilon_{xy} & \gamma_{yz}+\gamma_{zx}+\gamma_{xy}\\
    \end{pmatrix}.
\end{equation}
\\Near the $K$ point the latter can be expanded to the first order, due to the fact that the sum of the $\epsilon$ vanishes in that point:
\begin{equation}
    H_{{\rm{eff}}}=\begin{pmatrix}
    t_1 & \frac{1}{2}(t_2+2t_3)(\delta k_X-ik_Y)\\
    \frac{1}{2}(t_2+2t_3)(\delta k_X+ik_Y) & t_1
    \end{pmatrix},
\end{equation}
where we used the notation of $\delta k_X=(k_X-\frac{4\pi}{\sqrt{27}})$. We can write this Hamiltonian introducing the very compact notation of the Pauli spin matrices $\sigma_i$:
\begin{equation}
    H_{{\rm{eff}}}=\hbar v_F \hspace{0.1cm} \Vec{\delta k}\cdot \Vec{\sigma}+\rm{const.}
\end{equation}
where $\Vec{\delta k}$ is the $k$ vector centered around the K point, $\hbar v_F=\frac{1}{2}(t_2+2t_3)$ and we neglected the constants because we are interested in the qualitative behaviour. Using $t_2=0.04$ eV and $t_3=0.5$ eV we obtain $v_F\approx2.5\times10^{5}$ m s$^{-1}$ which is smaller than the graphene one ($v_F\sim10^6$~m~s$^{-1}$).
The eigenstates of this Hamiltonian are
\begin{equation}
    \ket{A_g\pm,\vek}=\frac{1}{\sqrt{2}}\left(\ket{a_g\text{Ti}_1,\vek}\pm e^{i\theta_{\vek}} \ket{a_g\text{Ti}_2,\vek}\right),
\end{equation}
where the phase $e^{i\theta_{\vek}}=\frac{\delta k_X\mp ik_Y}{|\Vec{\delta k}|}$, so that $\theta_{\vek}$ coincides with the geometrical angle between $\Vec{\delta k}$ and $k_X$.
The degeneracy near the $K$ point is repeated six times throughout the hexagonal Brillouin zone: however, these six points are not all equivalent. In particular, near the $K'$ point the Hamiltonian is the same with a change in sign in $\sigma_x$. If, for this second class of Dirac points, we reverse the ordering of the states in the matrix form of the Hamiltonian, we find that the form of the Hamiltonian is equal and opposite to the one near $K$. Thus, with the new ordering of states, we can write the effective Hamiltonian near $K'$ as
\begin{equation}
    H_{{\rm{eff}}}'=-\hbar v_F \hspace{0.1cm} \Vec{\delta k}\cdot \Vec{\sigma}+\rm{const.}
\end{equation}
The eigenstates of this Hamiltonian now become
\begin{equation}
    \ket{A_g'\pm,\vek}=\frac{1}{\sqrt{2}}\left(\ket{a_g\text{Ti}_2,\vek}\mp e^{i\theta_k} \ket{a_g\text{Ti}_1,\vek}\right).
\end{equation}
The degenerate points therefore group into three points equivalent to $K$ and three points equivalent to $K'$, forming two independent sublattices. This is again exactly what happens in graphene, where in the usual treatment a two-valued degree of freedom is introduced, generally named valley degree of freedom, corresponding to which class of Dirac point is under analysis. In the space of this new degree of freedom one can introduce new Pauli matrices, which we will denote by $\tau_i$, so that the full Hamiltonian can be written as
\begin{equation}
    H_{{\rm{eff}}}=\hbar v_F \hspace{0.1cm} \Vec{\delta k}\cdot \Vec{\sigma} \otimes \tau_z +\rm{const.}
\end{equation}
where the state is now a four-spinor.
\begin{figure*}[t!]
    \centering
    \includegraphics[width=0.30\textwidth]{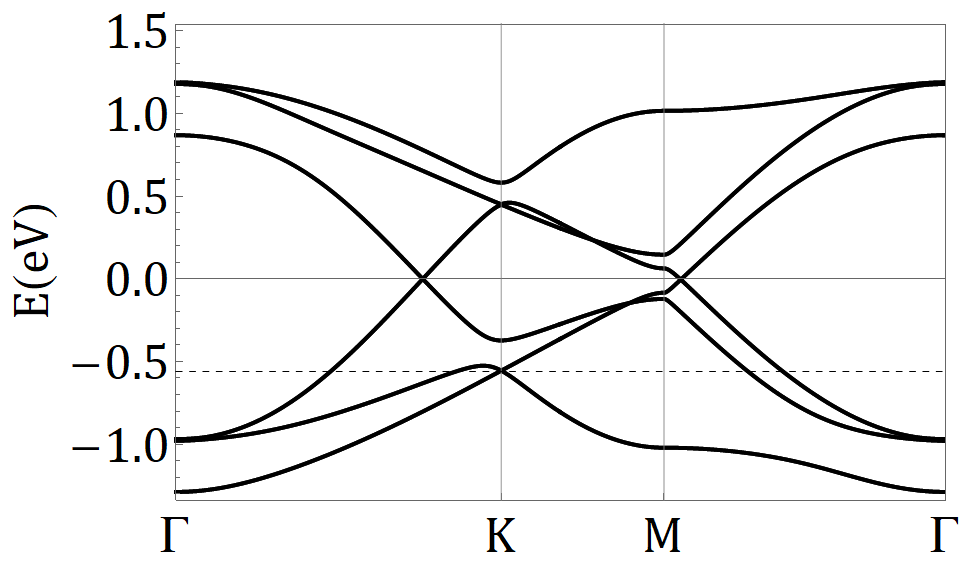}\hfill
    \includegraphics[width=0.30\textwidth]{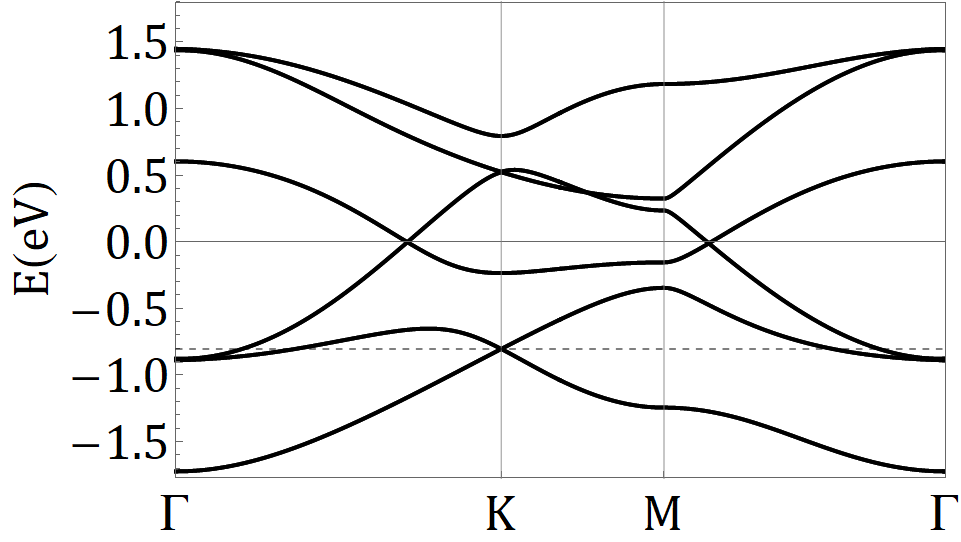}\hfill
    \includegraphics[width=0.30\textwidth]{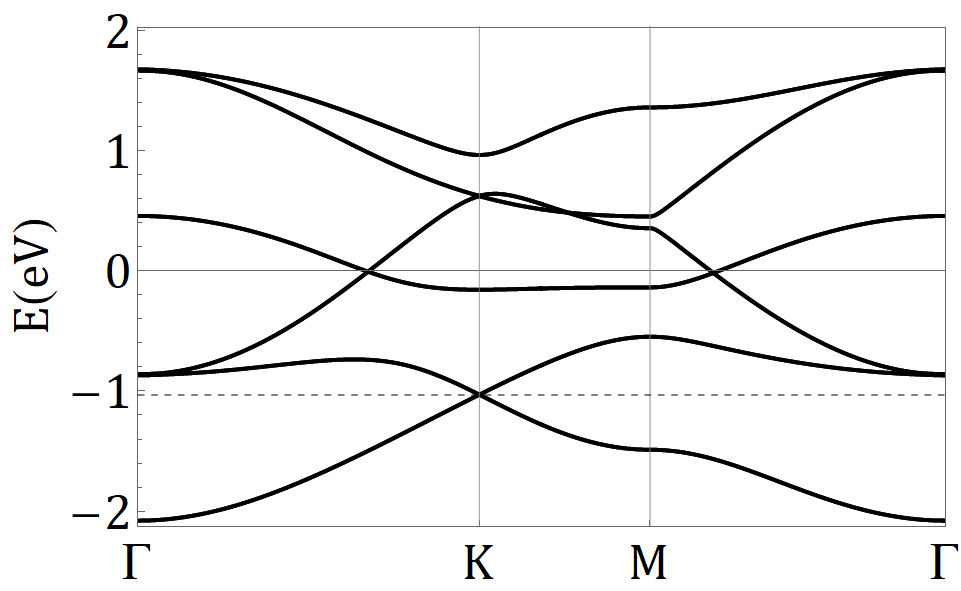}
    \caption{Evolution of the band structure for increasing strain applied to the surface. We show the band structure for a distortion angle increasing from $\delta=1^\circ$ (a), $\delta=3.2^\circ$ (b), $\delta=6^\circ$ (c). The dashed line corresponds to the filling needed to reach the intersection of the first two doublets at the $K$ point, namely $\mu=-0.56$ eV (a), $\mu=-0.815$ eV (b), and $\mu=-1.04$ eV (c). The electric field at the interface is fixed to $v=0$; the spin-orbit coupling is taken as $\lambda=0.01$ eV.}
    \label{Evolution}
\end{figure*}
We conclude that there is a resemblance between graphene's Dirac cone and (111) LAO/STO interface. An experimentally relevant difference is the fact that the Dirac point in the graphene stands exactly at the neutrality point of the system, between the conduction and valence band, while the Dirac point in the LAO/STO interface resides fully in the conduction band. Therefore, in order to investigate that point we need to fill sufficiently the $\ket{A_g}$ lower bands: the numerical predicted value (for our choice of parameters) is $n_{2D}=7.6\times10^{14}$ cm$^{-2}$. This value is comparable in order of magnitude with the typical charge carrier density predicted at these interfaces by the polar catastrophe model.
\\We now discuss the presence of the effective emergent coupling due to the SOC at the Dirac point. The energy splitting induced by the SOC has to be treated  carefully since, as we mention in the text, the mean value of $\Vec{L}$ is equal zero on the $a_g$ states, leading to no induced splitting to the first order in the SOC. Moreover, both the SOC and the TB terms are to be treated as small perturbations to the larger trigonal strain term. We explicitly verified that a gap among the $a_g$ states at the Dirac point only opens to the third order in perturbation theory: in particular the gap opening is well-described by the perturbative term in the Hamiltonian
\begin{equation}
   H_{Z}=\Delta_{SO}\sigma_Z\otimes\tau_z,
\end{equation}
with
\begin{equation}
    \Delta_{SO}=0.337 \frac{\lambda t_3^2}{\Delta^2}.
\end{equation}
In these formulas for simplicity we set $t_2=0$.
Therefore the complete Hamiltonian on the Dirac points is well described by:
\begin{equation}
    H_{{\rm{eff}}}=\hbar v_F \hspace{0.1cm} \Vec{\delta k}\cdot \Vec{\sigma} \otimes \tau_z +\Delta_{SO}\sigma_Z\otimes\tau_z +\rm{const.}
\end{equation}
\section{Strain influence on hopping parameters}\label{Strainandhopping}
In this section we will discuss the effect of the strain on the hopping parameters. We will demonstrate that, even if the trigonal strain does change the hopping parameters, mainly due to the change in the bond lengths, the regime of $|\Delta|/t_3>1$ described in the text can always be reached. As we discussed in the main text, a trigonal distortion is a rigid shift of one layer with respect to the other along the (111) direction. First of all, this shift preserves the relative magnitude of the hopping parameters in different directions: since the trigonal strain preserves the $C_{3}$ symmetry, it does not give rise to new types of TB couplings violating $C_3$, and therefore it only affects the numerical values of the parameters $t_1$, $t_2$, and $t_3$. The numerical change is induced by two factors: the different angle in the projection of the orbitals, and the different bond lengths in the two centers Slater-Koster (SK) overlap integrals
\begin{equation}
    E_{l m, l^{\prime} m^{\prime}}=\int d \Vec{r} \hspace{0.1 cm}\bar{\psi}_{l m}(\Vec{r}-\Vec{d}) V(\Vec{r}-\Vec{d}) \psi_{l^{\prime} m^{\prime}}(\Vec{r}),
    \label{SKintegrals}
\end{equation}
where $\psi_{l m}(\Vec{r})$ is the wave function of the electrons localized on $\Vec{r}$ and $V$ is the ionic potential which mediates the hopping between the two sites. We discuss the two dependencies in turn.
\\On the one hand, the cosine of the angle $\delta$ between the Ti-Ti joining line and the $z$ direction, which would be 1 in the equilibrium configuration (see Fig.~(\ref{angolo_strain})), changes in the distorted configuration. Since the (SK) integrals depend regularly on the cos($\delta$), this will affects the hopping parameters by powers of $\cos\delta\simeq1-\delta^2/2$, so that even for large angle of distortion, e.g. $10^\circ\sim0.17$ rad, the correction of the hopping parameters is of the order of 1.5\%. We will neglect the correction due to the distortion angle in comparison with the one due to the bond lengths.
\\On the other hand, the distances between two Ti atoms are changed by the distortion of the crystalline planes. The unperturbed angle between one side of the cubic lattice and the $(111)$ direction is $\theta_0\simeq54.7^\circ$. Under contraction of the crystalline planes, this angle increases to $\theta_0+\delta$.
The distance between two Ti atoms is decreased from $a_0$ to $a_0\sin\theta_0/\sin(\theta_0+\delta)$.
\begin{figure*}[t]
    \centering\scalebox{0.93}{
    \includegraphics[width=0.3\textwidth]{Z2Cont_v0_ban1.png}
    \hfill
    \includegraphics[width=0.30\textwidth]{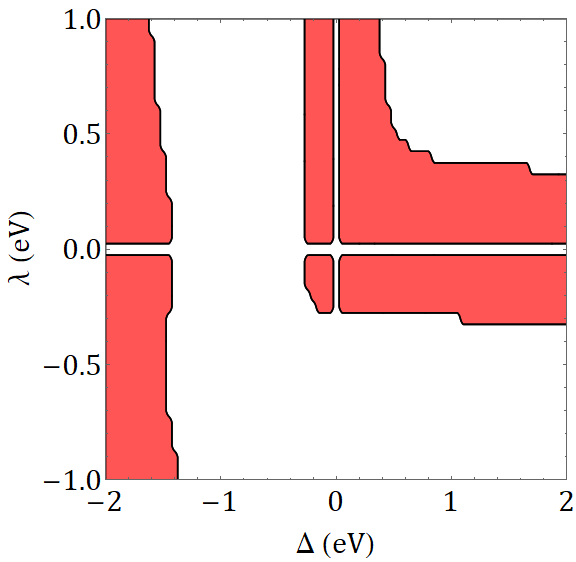}
    \hfill
    \includegraphics[width=0.30\textwidth]{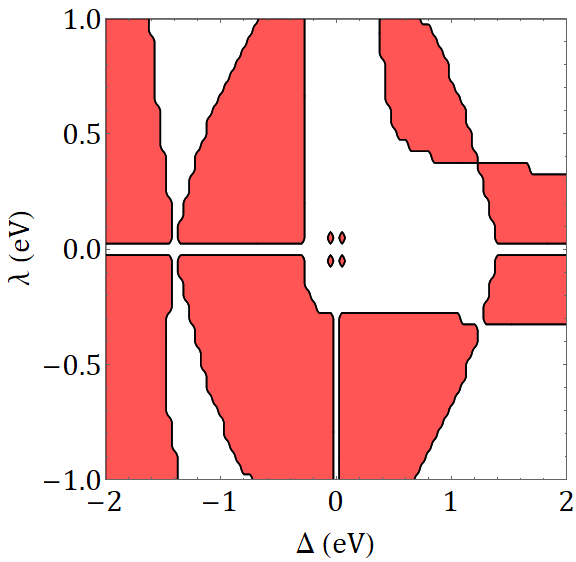}}
    \\\scalebox{0.93}{
    \includegraphics[width=0.30\textwidth]{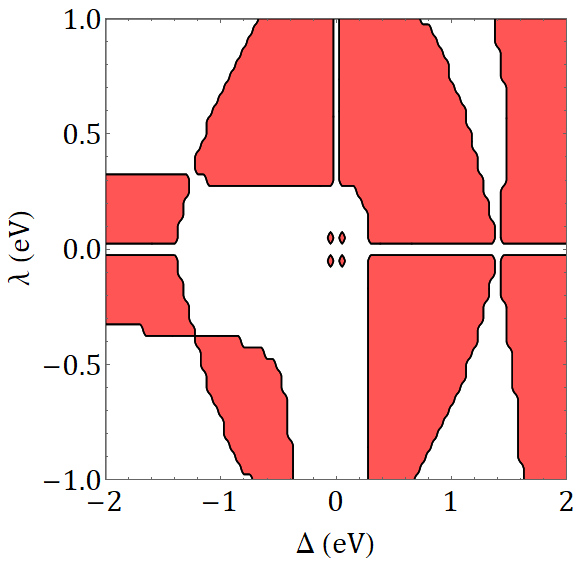}
    \hfill
    \includegraphics[width=0.30\textwidth]{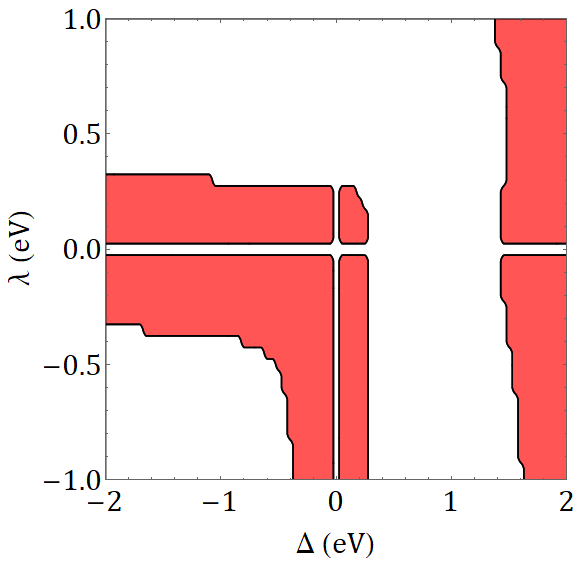}
    \hfill
    \includegraphics[width=0.30\textwidth]{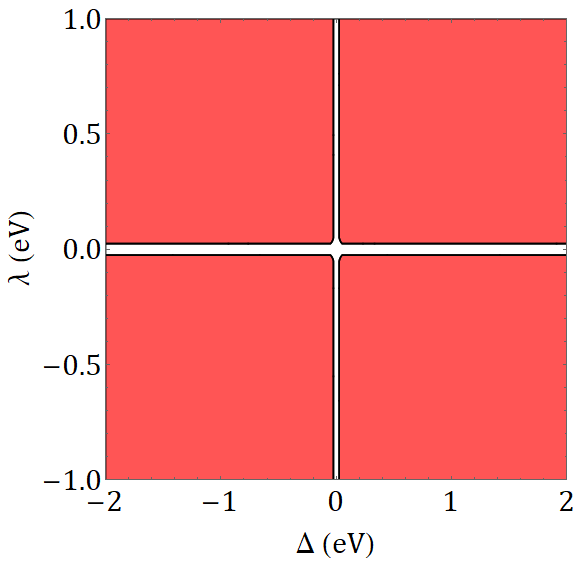}
    }
    \caption{Phase diagram for each of the six Kramers doublets in the $\lambda-\Delta$ plane with $v=0$ eV. The panels are ordered form lowest to the highest energy doublet. The red region corresponds to $Z_2=1$, while the white one corresponds to $Z_2=0$. Every diagram has been computed using an energy step of $0.05$ eV.}
        \label{cont1}
\end{figure*}
The SK integrals, and therefore the $t_3$ parameter, depend on the distance as $d^{-(l+l^{\prime}+1)}$ \cite{perroni2003modeling}, and since we are in the subspace of the $\{t_{2g}\}$ which have a quenched angular momentum, $l=1$, $l^\prime=1$ and thus $t_3\propto d^{-3}$. For a representative value $\delta=10^\circ$ (which is large compared to the angles needed to reach the regime $|\Delta|>t_3$), the hopping parameter $t_3$ changes from $0.5$ eV to $0.5\;\text{eV}/(0.9)^3\simeq 0.7$ eV. Therefore, due to the small values of the distortion angles needed to reach the regime of $|\Delta|/t_3>1$, the change in the hopping parameters is not expected to significantly influence our results. We emphasize in any case that our results are based on choices for the numerical parameters which are conservative: for example, for the constant $\kappa$ we have taken a value $\kappa=4.93$. The linearized muffin-tin orbitals approximation predicts larger values for $\kappa$, leading to $|\Delta|$ which are even larger, therefore requiring even smaller distortion angles to enter the regime of $|\Delta|/t_3>1$. In Fig.~(\ref{angolo_strain}) we show the dependence of $t_3$ and $\Delta$ as a function of the distortion angle. We can see that, since $t_3$ grows slower than $|\Delta|$, it is possible to reach the $|\Delta|/t_3>1$ regime already for $\delta\sim3^\circ$. In Fig.~(\ref{Evolution}) we show the evolution of the band structure for decreasing $\Delta,$ taking into account the correction to the hopping parameters due to the distortion angle $\delta$. The result shows that already for $\delta=6^{\circ}$ the Dirac cone is present, corresponding to $\Delta=-0.81$ eV.
\section{Topological phase diagrams}\label{AppendiceFigure}
\begin{figure*}[h!]
   \centering\scalebox{0.93}{
    \includegraphics[width=0.30\textwidth]{Z2Cont_l01_vsmall_ban1.png}
   \hfill
    \includegraphics[width=0.30\textwidth]{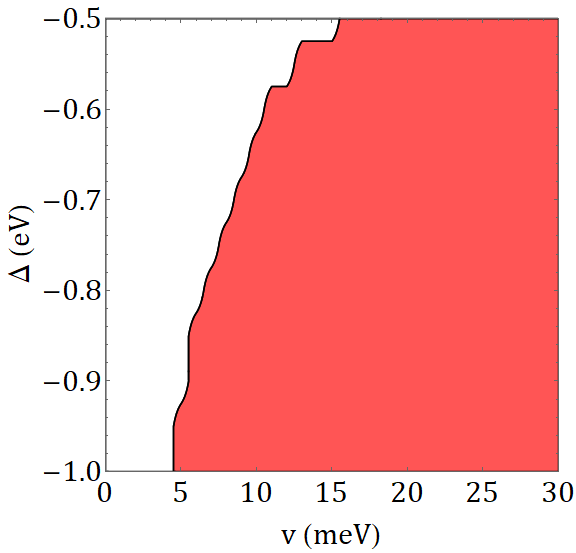}
    \hfill
   \includegraphics[width=0.30\textwidth]{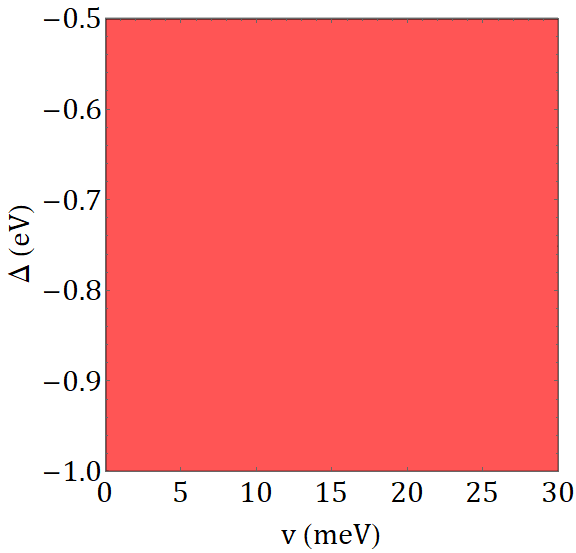}}
   \\\scalebox{0.93}{
   \includegraphics[width=0.30\textwidth]{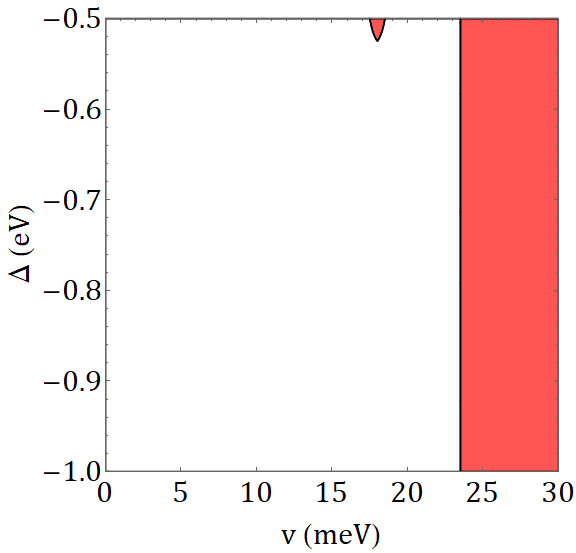}
    \hfill
    \includegraphics[width=0.30\textwidth]{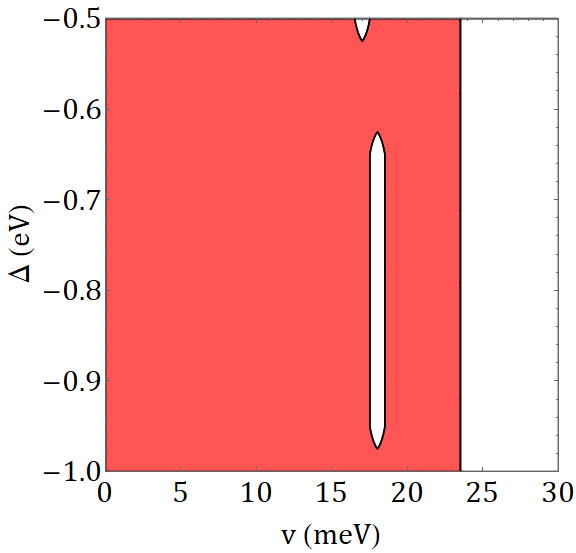}
    \hfill
    \includegraphics[width=0.30\textwidth]{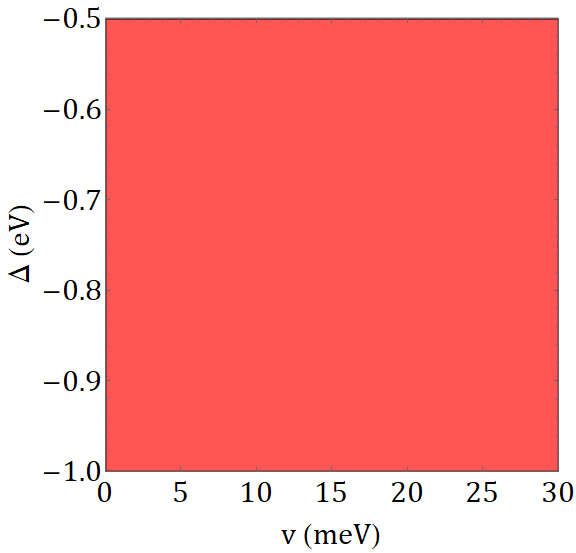}}
    \caption{Phase diagram for each of the six Kramers doublets in the $v-\Delta$ plane fixing $\lambda=0.1$ eV for small values of $v$ in the topologically interesting region. The panels are ordered form lowest to the highest energy doublet. The red region corresponds to $Z_2=1$, while the white one corresponds to $Z_2=0$. Every diagram has been computed using an energy step for $\Delta$ ($v$) of $0.05$ ($0.001$) eV.}
        \label{cont2}
\end{figure*}

\begin{figure*}
    \centering\scalebox{0.93}{
    \includegraphics[width=0.30\textwidth]{Z2Cont_D1_vsmall_ban1.png}
    \hfill
    \includegraphics[width=0.30\textwidth]{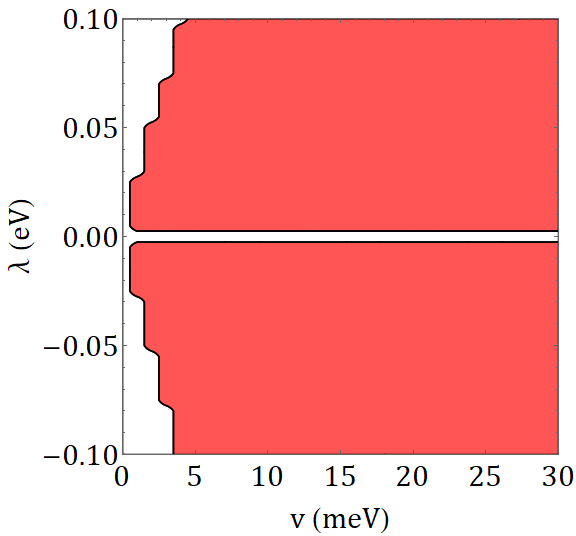}
    \hfill
    \includegraphics[width=0.30\textwidth]{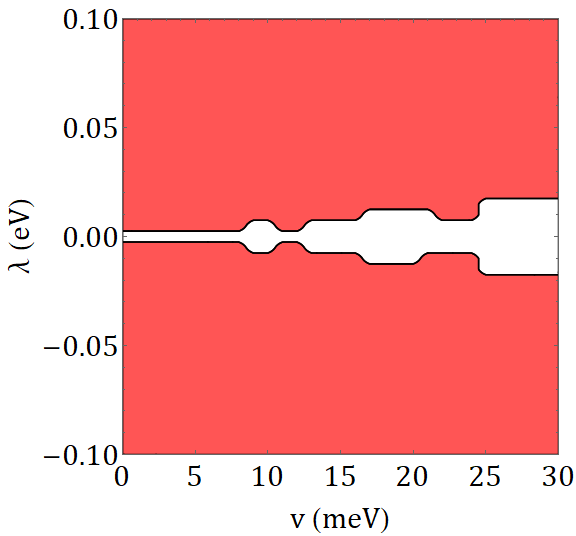}}\\
    \scalebox{0.93}{
    \includegraphics[width=0.30\textwidth]{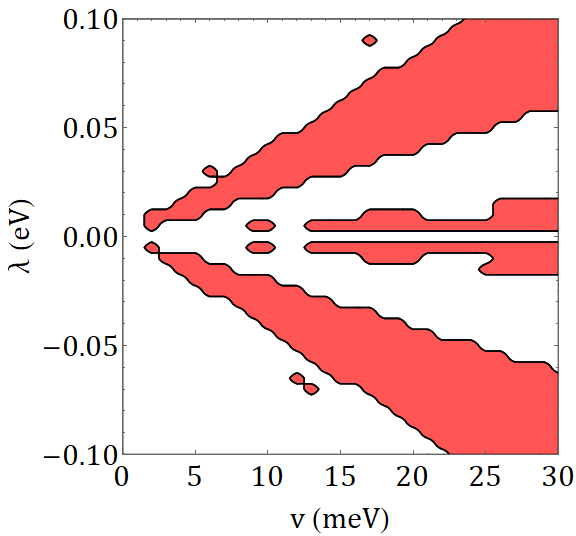}
    \hfill
    \includegraphics[width=0.30\textwidth]{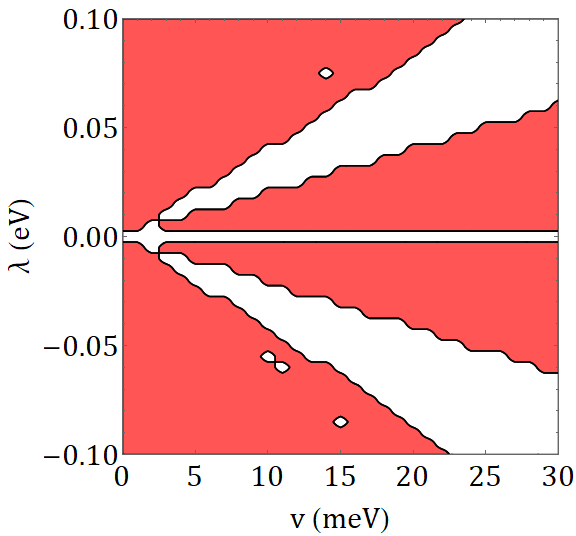}
    \hfill
   \includegraphics[width=0.30\textwidth]{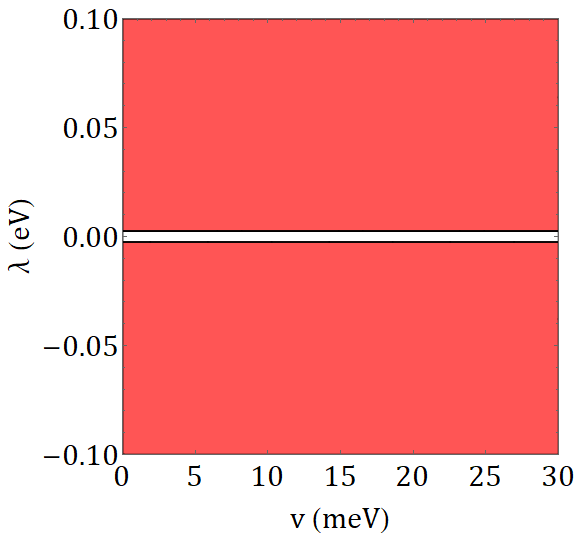}
   }
    \caption{Phase diagram for each of the six Kramers doublets in the $\lambda-v$ plane fixing $\Delta=-1$ eV. The panels are ordered form lowest to the highest energy doublet. The red region corresponds to $Z_2=1$, while the white one corresponds to $Z_2=0$. Every diagram has been computed using an energy step for $\lambda$ ($v$) of $0.005$ ($0.001$) eV.}
    \label{cont3}
\end{figure*}

In this Appendix we show the topological phase diagrams for all the six Kramers doublets. In Figs.~(\ref{cont1}), (\ref{cont2}) and (\ref{cont3}) we show respectively the $Z_2$ behaviour in the the $\Delta-\lambda$ plane, in the $v-\lambda$ plane and in the $v-\Delta$ plane.
\clearpage
\bibliographystyle{unsrt}
\bibliography{Bib.bib}

\end{document}